\documentclass[PAPER, atlasdraft=false, UKenglish, cernpreprint, texmf, orcidlogo]{atlasdoc}
\usepackage{atlaspackage}
\usepackage{atlasbiblatex}

\usepackage{atlasphysics}
\usepackage[output=true, shift=true]{}
\graphicspath{{logos/}{figures/}}

\usepackage{ANA-JETM-2022-06-PAPER-defs}
\usepackage{PMGRefs-defs}
\usepackage{mydocument-defs}
\usepackage{atlasjetetmiss}

\AtlasTitle{A precise measurement of the jet energy scale \\ derived from single-particle measurements and \\ in situ techniques in proton--proton collisions at $\sqrt{s}=$ 13 TeV with the ATLAS detector}

\AtlasAbstract{%
The jet energy calibration and its uncertainties are derived from measurements of the calorimeter response to single particles
in both data and Monte Carlo simulation using proton--proton collisions at $\sqrt{s} = 13$~TeV collected with the ATLAS detector during Run~2 at the Large Hadron Collider.
The jet calibration
uncertainty for anti-$k_T$ jets with a jet radius parameter of \textit{R}$_{\rm jet} = 0.4$ and in the central jet rapidity region

Excellent agreement is found with earlier determinations obtained from $p_{\rm T}$-balance based in situ methods ($Z/\gamma$+jets).
The combination of these two independent methods results in the most precise jet energy measurement achieved so far with the ATLAS detector with a relative uncertainty of $0.3\%$ at $p_{\rm T}=300$~GeV and $0.6\%$ at $4$~TeV.

The jet energy calibration is also derived with the single-particle calorimeter response measurements separately for quark- and gluon-induced jets
and furthermore for jets with \textit{R}$_{\rm jet}$ varying from $0.2$ to $1.0$ retaining the correlations between these measurements.
Differences between inclusive jets and jets from boosted top-quark decays, with and without grooming the
soft jet constituents, are also studied.
}
\PreprintIdNumber{CERN-EP-2024-195}

\AtlasJournalRef{\EPJC 85 (2025) 927}
\AtlasDOI{10.1140/epjc/s10052-025-14409-1}

\hypersetup{pdftitle={ATLAS document},pdfauthor={The ATLAS Collaboration}}

\begin{document}

\maketitle

\section{Introduction}
%

Jets, collimated sprays of hadrons, are ubiquitously produced in proton--proton ($pp$) collisions at the Large Hadron collider (LHC).
In the theory of the strong interaction, Quantum Chromodynamics (QCD), they are interpreted as the footprint of quarks and gluons produced at high transverse momentum (\pt). The measurement of the energy of jets plays an important role for all jet-based measurements at the ATLAS experiment, and the uncertainty in the jet energy calibration often limits the precision of important physics results.

In ATLAS, the jet energy measurement~\cite{PERF-2014-02, PERF-2015-09, PERF-2016-04, JETM-2018-05, ATLAS:2023tyv}
relies on corrections based on Monte Carlo (MC) simulations
modelling the detector response and by assessing \insitu{} how well the MC simulations describe data.
The calibration of the average jet \pt is called the `jet energy scale' (\JES)
and consists of several steps.
The first correction based on MC simulations ensures that the reconstructed jet energy
equals, on average, the energy of the corresponding jet formed by stable particles interacting with the ATLAS detector.
Several effects are corrected including the non-compensating nature of the ATLAS calorimeter having a lower response to hadrons than to electrons or photons, energy deposits outside jets and energy losses in dead material before and in between the calorimeters.
The jet response is also equalized as function of variables, like, e.g. the energy fraction in the hadronic calorimeter or the number of tracks, that are sensitive to jet fragmentation. In this procedure the average jet response is not changed, but the jet resolution is improved.
In a final step of the jet calibration, the detector response to the jet in the MC simulation is compared to the one in the data
by exploiting the \pt-balance in certain event topologies where the jet recoils against a well-measured object like a photon or a $Z$ boson decaying into electron or muon pairs ($Z/\gamma$+jet).

The jets are first calibrated in the central region of the detector.
The \pt-balance in dijet events with a central and a forward jet extends this calibration to more forward regions.
At higher \pt{}, the uncertainties associated with this extrapolation tend to be sub-dominant such that the primary
motivation to improve the precision of the jet energy calibration with independent methods
is in the central region of the detector. In this paper, the focus is therefore on the most central pseudo-rapidity region $|\eta|<0.8$.\footnote{\AtlasCoordFootnote}

In the standard jet calibration methods, the main systematic uncertainties are related to the
knowledge of the \pt-balance in $Z/\gamma$+jet event topologies in data and MC simulation.
Since jets are composed of collections of single particles, the single-particle response measurements
can be used to derive the \JES{} and the corresponding \JES{} uncertainties. During the beginning and end of the LHC \RunOne, ATLAS derived the data-to-MC jet response correction using calorimeter response measurements of single particles from $pp$ collisions at $\sqrt{s}= 7$ and $\sqrt{s}=8$~\TeV{} \cite{PERF-2011-05,PERF-2015-05} and relied on test-beam measurements for particles in the high-\pt{} regime.
The data-to-MC response shift was compatible with the one derived from the \pt-balance based  methods, but the resulting  uncertainty was larger.
Therefore, the single-particle-based uncertainty was only used for very high-\pt{} jets beyond the kinematic reach of the \pt-balance based methods.
A similar method was also used for the determination of the $\tau$-lepton energy scale for hadronic $\tau$-lepton decays at a collision energy of $8$~\TeV{} \cite{PERF-2013-06}.

In  \RunOne{}, the single-hadron response measurements, i.e., the ratio of the calorimeter energy to the momentum of an isolated track (\EoP), were only performed in minimum bias collisions.
The kinematic reach was limited to a track \pt{} of $20-30$ \GeV, since high-\pt{} tracks are preferentially found within jets and are not isolated.
The higher \pt{} range was therefore covered with pre-data-taking test-beam measurements
where a complete slice of the ATLAS detector was exposed to pion beams~\cite{Abat:2010zza}. Differences between particle types were studied with MC simulations.

In \RunTwo, where the LHC collided protons on protons at a centre-of-mass energy of $13$~\TeV, sufficiently large data sets were available to exploit isolated single pions resulting from the hadronic decay of $\tau$-leptons produced in $W$ boson decays (\Wtaunu).
The \EoP{} response for isolated pions from $\tau$-lepton decays was measured for $10< \pt <300$ \GeV{} with small uncertainties (below $1.5\%$) \cite{JETM-2018-08}. This much larger kinematic reach and the small associated uncertainty opened up a new level of precision for the \JES.

This paper presents the \JES{} and the corresponding uncertainty derived from single-particle response measurements and MC simulation studies.
It is derived for \AKT-jets \cite{Cacciari:2008gp} with a nominal radius parameter $\Rjet = 0.4$, and clustered using the
\Fastjet{} package \cite{Fastjet,Cacciari:2005hq}. Only the central pseudo-rapidity region $|\eta|<0.8$ is used.
The other calorimeter regions can be calibrated with the \insitu{} dijet pseudo-rapidity intercalibration method
where the forward jet is balanced by a central jet~\cite{PERF-2014-02, PERF-2015-09, PERF-2016-04, JETM-2018-05, ATLAS:2023tyv}. The single-particle response method also allows for the derivation of the \JES{} calibration factors and the uncertainties for quark- and gluon-induced jets, and for different jet definitions such as for varying the chosen value of \Rjet{} parameters.

The jets are formed from topological clusters of calorimeter cells \cite{PERF-2014-07} on the electromagnetic scale\footnote{The electromagnetic scale is the calibration of the ATLAS calorimeter that ensures that the energy deposited by electromagnetic showers is correctly measured.} (\EMTopo) and are corrected with a MC-based calibration \cite{PERF-2014-02,JETM-2018-05,PERF-2012-01,PERF-2011-03}.
The \JES{} derived from single-particle measurements is compared to that obtained from the \insitu{} \pt-balance based methods, and subsequently, combined with it.
Since the uncertainties of the two methods are largely independent and of somewhat similar amplitude the combination gives significant improvements.
The combination for particle-flow (\PFlow) jets~\cite{PERF-2015-09}, where the information from the tracking and calorimeter
system is combined to achieve optimal performance, is only performed in a restricted jet \pt{} range.

The paper is structured as follows:
Section~\ref{sec:mc} briefly describes the ATLAS detector and the MC simulation.
The method to derive the \JES{} from single-particle response measurements (deconvolution method)
is described in Section~\ref{sec:method}. The input measurements and their systematic uncertainties are explained.
The final results of the \JES{} together with a detailed break-down of the uncertainties are presented
in Section~\ref{sec:result} for quark- and gluon-initiated jets separately. The results are
compared to the ones obtained with the $Z/\gamma$+jets methods.
Section~\ref{sec:combination} presents the combination of the \JES{} obtained with the deconvolution method
and with the  $Z/\gamma$+jets method.
The new uncertainties are compared to the ones
previously available for physics analyses in Section~\ref{sec:comparison}.

In Section~\ref{sec:rscan} the \JES{} is obtained for jets with various radius parameters scanned
from $0.2$ to $1.0$. Differences between inclusive jets produced by strongly interacting particles and jets from boosted top quarks
with and without grooming soft jet constituents are studied. Finally, the conclusions are given in Section~\ref{sec:conclusion}.


%
%
%
%
%
%
%

%
\pagebreak
\section{ATLAS detector and Monte Carlo simulation}
\label{sec:mc}

The ATLAS experiment~\cite{PERF-2007-01} at the LHC is a multipurpose particle detector
with a forward--backward symmetric cylindrical geometry and a near \(4\pi\) coverage in
solid angle. %
It consists of an inner tracking detector surrounded by a thin superconducting solenoid
providing a \qty{2}{\tesla} axial magnetic field, electromagnetic and hadronic calorimeters, and a muon spectrometer.
The inner tracking detector covers the pseudo-rapidity range \(|\eta| < 2.5\).
It consists of silicon pixel, silicon microstrip, and transition radiation tracking detectors.
Lead/liquid-argon (LAr) sampling calorimeters provide electromagnetic (EM) energy measurements
with high granularity within the region \(|\eta|< 3.2\).
A steel/scintillator-tile hadronic calorimeter covers the central pseudo-rapidity range (\(|\eta| < 1.7\)).
The endcap and forward regions are instrumented with LAr calorimeters
for EM and hadronic energy measurements up to \(|\eta| = 4.9\).
The muon spectrometer surrounds the calorimeters and is based on
three large superconducting air-core toroidal magnets with eight coils each.
The field integral of the toroids ranges between \num{2.0} and \qty{6.0}{\tesla\metre}
across most of the detector.
The muon spectrometer includes a system of precision tracking chambers up to \(|\eta| = 2.7\) and fast detectors for triggering up to \(|\eta| = 2.4\).
The luminosity is measured mainly by the LUCID--2 detector located close to the beam-pipe.
A two-level trigger system is used to select events~\cite{TRIG-2016-01}.
The first-level trigger is implemented in hardware and uses a subset of the detector information to accept events at a rate below \qty{100}{\kHz}.
This is followed by a software-based trigger that
reduces the accepted event rate to \qty{1}{\kHz} on average,
depending on the data-taking conditions in the LHC \RunTwo.

A software suite~\cite{SOFT-2022-02} is used in data simulation, in the reconstruction
and analysis of real and simulated data, in detector operations, and in the trigger and data acquisition
systems of the experiment.
The ATLAS detector is modelled with a detailed simulation~\cite{SOFT-2010-01} based on \GEANT~\cite{Agostinelli:2002hh}.
The hadronic shower simulations mainly consist of three kinematic regions:
physics interactions of the nucleus for momenta of the incident particles around $1$~\GeV{} (pre-compound model),
nucleon-nucleon cascades in the nucleus in the medium \pt{} region,
and the modelling of the nuclear break-up for high \pt.
The choice of the models as function of the kinematic regime is called the `physics list' in the \GEANT simulation.
The ATLAS default simulation is based on the \texttt{FTFP\_BERT\_ATL}~\cite{physicsList} physics list. In this physics list the \textit{Bertini cascade model}~\cite{Guthrie:1968ue} is used for hadrons with energy below $12$~\GeV, and the \textit{Fritiof string model}~\cite{ANDERSSON1987289,NILSSONALMQVIST1987387} to model the nuclear break-up for hadrons with energy above 9~\GeV. The probability of using each model changes smoothly across the region of overlap.

As alternative to the Bertini cascade model and the Fritiof model, the \textit{quark gluon string model} (QGSP)~\cite{folger2003string} and
the \textit{binary cascade model}~\cite{binary} are used for systematic studies.
Other variations include a high-precision tracking of electromagnetic showers and very low energy neutrons,
a different set of elastic $pp$ cross-sections and a different treatment of diffractive processes.
Further systematic uncertainties are estimated by using simulations with varied detector geometries.
The material in the inner detector is increased in specific regions and overall by $5\%$ and material is added between the
electromagnetic and the hadronic calorimeters. Moreover, the detector geometry is distorted in the inner detector end plate,
the cryostat and before and after the calorimeter presampler. These variations are constructed based on the degree of knowledge of the detector geometry based on measurements taken during construction and studies of interactions in the inner detector~\cite{PERF-2015-07}.

The nominal MC simulation of a sample of inclusive dijet events is the same as used in the calibration of jets in ATLAS~\cite{PERF-2014-02, PERF-2015-09, PERF-2016-04, JETM-2018-05, ATLAS:2023tyv}.
It is produced using \PYTHIA[8.235]~\cite{Sjostrand:2014zea} with the A14 set of tuned parameters~\cite{ATL-PHYS-PUB-2014-021}.  \PYTHIA{} uses a leading-order $2\rightarrow 2$ matrix element interfaced with a parton distribution function (PDF) to model the hard process. The PDF used is \NNPDF[2.3lo]~\cite{Ball:2012cx}. Additional radiation is modelled in the leading-logarithm approximation using \pT-ordered parton showers. Multiple parton-parton interactions and underlying event are also simulated. Hadronisation is based on the Lund string model~\cite{Sjostrand:1984ic,Andersson:1983ia}, and \EVTGEN[1.6.0]~\cite{Lange:2001uf} is used to decay heavy hadrons.

An alternative sample of inclusive dijet events is generated with
\sherpa 2.2.11 \cite{,Bothmann:2019yzt} with the {\textsc CT14nnlo} PDF~\cite{Dulat:2015mca}. A leading-order $2\rightarrow 2$ matrix element is interfaced with a parton shower using Catani--Seymour dipole factorisation~\cite{Schumann:2007mg}.
The cluster based hadronisation module \ahadic{}~\cite{Winter:2003tt} is used
with re-tuned parameters that improve agreement with the LEP data of the fractions of different hadrons within hadronic jets~\cite{Chahal:2022rid}. The internal \textsc{Hadrons++} module is used to decay heavy hadrons.

The \JES{} of large-\Rjet{} jets, where boosted top quarks and $W$ bosons are merged in one large jet,
is studied using a sample with new hypothetical heavy $Z$ bosons (\Zprime) that further decay into top quark pairs.
This sample is simulated using \Pythia[8.212] with the A14 tune
and the leading-order \NNPDF[2.3lo] PDFs.
The cross-section of the hard-scattering process and the branching fractions were modified
such that a fairly flat jet \pt{} spectrum is obtained between 250~\GeV\ and 1.5~\TeV\ for
$b$-jets, $c$-jets, and light-flavour jets, with a falling tail of the \pt{} distribution
populated to 3~\TeV\ for each flavour.

Furthermore, simulated inclusive inelastic $pp$ collisions were overlaid to model additional pile-up collisions in the same and neighbouring bunch crossings. These were generated with \PYTHIA[8.186]~\cite{Sjostrand:2007gs}
using the \NNPDF[2.3lo] PDFs and the A3 set of tuned parameters~\cite{ATL-PHYS-PUB-2016-017}.


%
%
%
\pagebreak
\section{Deconvolution method and input measurements}
\label{sec:method}

\begin{figure}[htpb]
\centering
\includegraphics[width=0.9\linewidth]{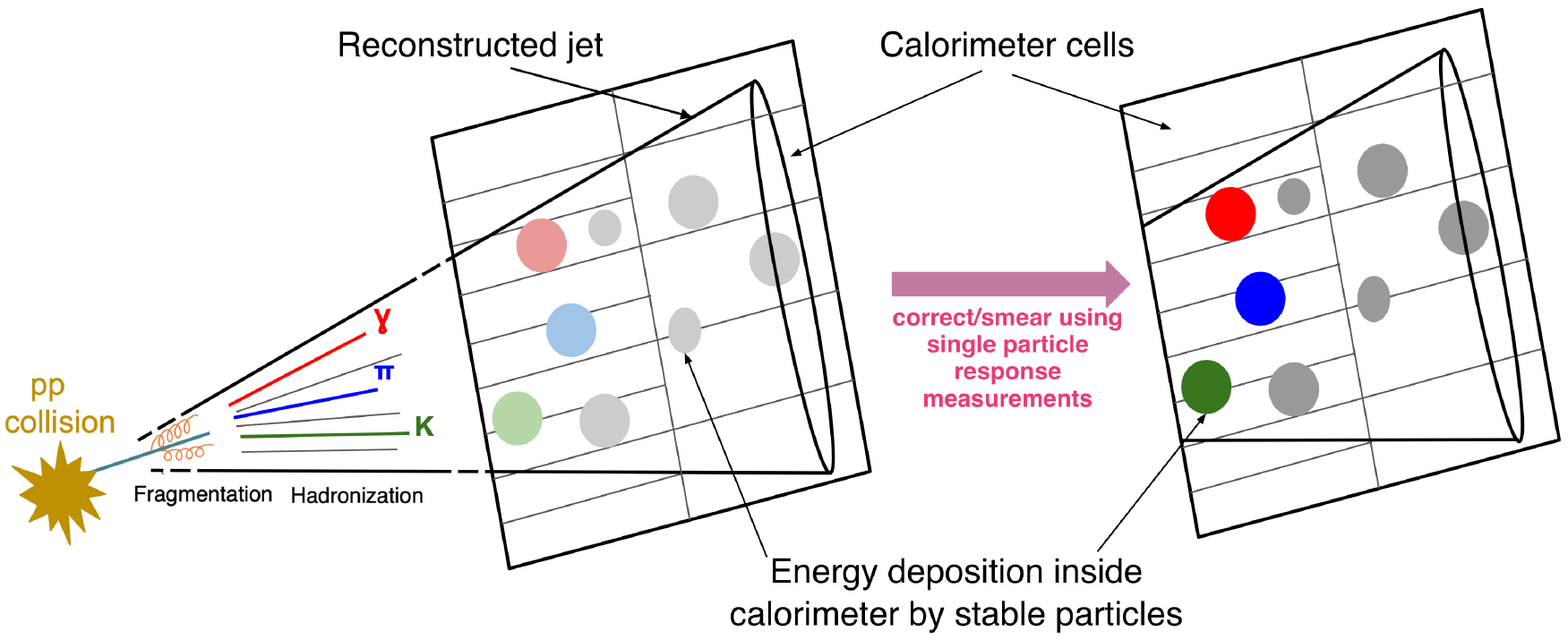}
\vspace{-2.cm}
\caption{Sketch of the particles associated with a jet and the corresponding energy depositions in the calorimeter. The colors represent the energy depositions of the various particle types. In the deconvolution method, the energy depositions are corrected and smeared to obtain the data-to-MC jet energy correction from calorimeter response measurements to single particles.}
\label{fig:EoP-diagram}
\end{figure}

The jet energy calibration corrects the measured jet energy for several effects like calorimeter non-compensation, energy losses in the dead material or energy deposited outside the jet. In a first step it is derived from MC simulation comparing the reconstructed and the true jet energy.\footnote{
The true jet energy is derived from the stable particles associated with the jet with a proper lifetime $c \tau > 10$ mm (excluding neutrinos and muons) impinging the ATLAS detector as given by the MC simulation before any detector simulation.}
In a second step, the calorimeter response to jets
in data and MC is evaluated, and the data-to-MC ratio is used as the jet energy response correction.
In the deconvolution method~\cite{PERF-2011-05,PERF-2015-05}, the jet response correction is determined by viewing the reconstructed jet as a superposition of energy depositions from the single particles constituting the jet. This method is based on MC simulation, but uses single-particle calorimeter response measurements for various particle types in data and MC to compute the data-to-MC jet response correction.
A schematic overview of the method is shown in Figure~\ref{fig:EoP-diagram}.

In the ATLAS MC simulation, the energy depositions produced in the ATLAS detector by each individual particle, called `truth' particle in the following, can be identified. Therefore for each energy deposition in the calorimeter
the kinematics and the particle type (electron, photons, pions, protons etc.) are known.
The energy depositions of the particles associated to jets can be used to calculate data-to-MC jet response correction
using measurements of the calorimeter response to single particles in data and MC simulation.
For each calorimeter cell associated to the reconstructed jet, the energy deposits of the truth particles are
identified and corrected by the measured ratio between data and MC of the
calorimeter response to an isolated particle of the corresponding particle type with the same kinematics.

The ratio of the total energy depositions within the jet after and before the correction gives the
data-to-MC jet response correction (${\rm R}_{\rm Data} /{\rm R}_{\rm MC}$).
The mean of this jet response correction is used for jet calibration.
The data-to-MC jet energy correction uncertainty is calculated from the single-particle response uncertainties. Each uncertainty component in the single-particle response gives an independent uncertainty component for the \JES. The numerical evaluation of the uncertainty in the \JES{} is performed with MC pseudo-experiments. In each pseudo-experiment, the jet energy scale correction is calculated after randomly varying the MC single-particle energy response within the appropriate uncertainty range, as given by the measured single-particle data-to-MC ratio. Each particle is treated independently. The final \JES{} uncertainty is then derived from the spread of the distribution of the jet energy scale over all pseudo-experiments.

The sequence of the randomly changed parameters are kept fixed within each pseudo-experiment
such that the energy response correlations are properly taken into account.
This method is chosen as different truth jets have different spectra of truth particle energies and types, and with this method the average corrections for all jets as a function of \pt{} can be obtained.

The measured data-to-MC single-particle energy shifts and the corresponding uncertainties
depend on the particle type and the kinematic range.
They are obtained from measurements of the single-particle response in data and MC simulation.
The calorimeter response to single charged hadrons
\EoP{} is defined as the ratio of the average energy deposited by an isolated charged particle
in the calorimeter ($E$) to the momentum of its inner detector track ($p$).

The \EoP{} measurement for single pions with $10< \pt < 300$~\GeV{} is performed using $W$ bosons decaying into $\tau$-leptons where the $\tau$-lepton decays into a pion.
The $W$ bosons are produced by the $W$+jets process or in $t\bar{t}$ events from $pp$ collisions at $13$~\TeV{} using the full \RunTwo{} data sets corresponding to an integrated luminosity of $139$~{\rm fb}$^{-1}$.
A missing transverse momentum trigger is used. The $\tau$-leptons are selected as isolated tracks
which predominantly selects one-prong decays. The effect of the neutral pions that can be produced in $\tau$-lepton decays is substracted on a statistical basis.
More details can be found in Ref.~\cite{JETM-2018-08}.
The measurements of single charged hadrons for $\pt<10$~\GeV{} were performed following the methodology of Ref.~\cite{PERF-2015-05}
using a minimum bias data sample obtained in $pp$ collisions at $13$~\TeV{}
from a special dedicated low pile-up run where an average of two additional collisions are expected.
Both of the measurements use data sets processed employing the same thresholds
for pile-up noise suppression corresponding to \RunTwo{} pile-up conditions.
The measurement of the electron and photon energy scale and uncertainty is performed using a tag-and probe technique in a data set with $Z$ bosons decaying into electron pairs~\cite{EGAM-2018-01,EGAM-2021-02}.
It also uses the full \RunTwo{} $13$~\TeV{} data set corresponding to an integrated luminosity of $139$~{\rm fb}$^{-1}$.

For all hadron types the same data-to-MC corrections are used.
To account for differences of the calorimeter response to hadrons of other types,
additional systematic uncertainties are estimated from MC simulation or previous measurements.
Jet fragmentation effects are addressed by comparing the results obtained with the \Pythia{} MC event generator  with the ones from \Sherpa.

The kinematic ranges of the single-particle measurements and the uncertainty estimations are summarised in Table~\ref{table:deconvolutioninput}
and explained in more detail in the following.

%
\begin{table}
\centering
\caption{
Overview of the input measurements and uncertainties used by the deconvolution method used for the calculation of the data-to-MC jet response correction.}
\begin{tabular}{cccc}
\hline
Particle & \pt[{\rm GeV}] & comment \\
\hline
hadrons & $\pt < 0.5$          & $5$ \% uncertainty \\
& $0.5<\pt <10$        & isolated tracks in minimum bias collisions \\
& $10 < \pt <300$      & pions from $\tau$-lepton decays in \Wtaunu{} sample \\
& $\pt > 300$          & MC out-of-range studies and using JES uncertainty constraint \\ \hline
$e/\gamma$ & $5< \pt < 200$    & standard $e/\gamma$ calibration from $Z \to e^+ e^- $  \\
&                      & and $0.5 \%$ uncertainty for effects in jets \\
& $\pt > 200$          & extrapolation based on electron calibration \\
& $\pt < 5 $           & additional $1\%$ uncertainty from $\pi^0 \to \gamma \gamma$ mass peak constraint \\
\hline
\end{tabular}
\label{table:deconvolutioninput}
\end{table}
%


\subsection*{Systematic uncertainties in the \EoP{} measurement in the \Wtaunu{} sample}
The statistical uncertainty in the \EoP{} measurements in the \Wtaunu{} sample is treated
as a systematic uncertainty in the \JES.
Possible variations of the background relative to the single-pion signal from $\tau$-lepton decays, i.e.
muons, electrons or jets with extreme fragmentation patterns, are estimated by varying selection criteria.
The peak of the \EoP{} distribution is determined using a fit to a Gaussian distribution and an empirically found Landau
distribution to account for background events. The systematic uncertainties are estimated from the
fit closure in a MC sample and by varying the shape of the background distribution using different MC generators.
Moreover, the simulated response of the background from additional $\pi^0$ in $\tau$-lepton decays is considered. %

The energy deposition from pile-up events is subtracted on a statistical basis using
the same technique as used for jets based on the event \pt-density ($\rho$)~\cite{Cacciari:2007fd}.
The non-closure of the pile-up correction and possible mis-modelling of pile-up effects
is estimated by splitting the data sample into two samples with an average of low ($20$) and high ($50$) number of pile-up events.

Biases in the track momentum measurement are addressed by modifying the event selection
such that the isolated track is matched to a muon allowing for a cross-calibration of the track \pt{} to the
muon \pt{} that is well measured using $Z$ bosons decaying into muon pairs~\cite{MUON-2022-01}.
The statistical uncertainty as well as the muon track calibration uncertainties are considered.
Effects from the muon track resolution which depend on the underlying \pt{} spectrum are evaluated as well.
Further details on the relative sizes of these components as a function of \pt{} can be found in Ref.~\cite{JETM-2018-08}.
A flat uncertainty of $0.15$\% is used that is also in line with the studies on the sagitta
bias using $J/\Psi$ and $Z$ boson decays into muon pairs~\cite{IDTR-2019-05}.

\subsection*{Systematic uncertainties in the \EoP{} measurement in the minimum bias sample}

The largest systematic uncertainty in the \EoP{} measurement in the minimum bias sample is due to the cluster seeding probability, which gives the fraction of charged isolated hadrons that deposit energy in the calorimeter that is too low compared with the noise thresholds.
The ratio of the cluster seeding probability
in data and MC is taken as a systematic uncertainty.
This uncertainty is about $30\%$ up to about $1$~\GeV, $10\%$ between $2$ and $3$~\GeV{}
and it is negligible at $10$~\GeV.
Since the cluster seeding probability might be different for isolated single particles
or particles within jets, the full effect is treated as an uncertainty and not as a jet response correction.

As a cross-check the energy fraction of truth particles depositing no energy in the calorimeter was also evaluated in the MC simulation.
If the total amount of this energy fraction is taken as uncertainty, this method gives as expected a somewhat smaller uncertainty
as for example it does not include effects from the noise thresholds of the topological clusters.

An additional $5\%$ uncertainty is taken for charged hadrons
below the measured range, i.e.  $\pt<0.5$ \GeV{} as estimated in Ref.~\cite{PERF-2015-05}. Even with this large uncertainty
the energy depositions of such low-\pt{} hadrons have a negligible effect,
since they amount to only a tiny fraction of the jet energy.
The statistical uncertainty in the \EoP{} measurement is also considered.
The same uncertainty in the track momentum measurements as in the \Wtaunu{} \EoP{} measurement is used.
Effects from pile-up collisions are estimated by comparing the \EoP{} measurement
in events with one and two additional collisions.
The influence of neutral particles close to the isolated track used in the \EoP{} measurements is estimated and subtracted
using a data set where the energy deposit in the
electromagnetic calorimeter is compatible with a minimum ionising particle~\cite{PERF-2015-05}.
This uncertainty is estimated to be $1\%$ ($3\%$) for $p>2$~\GeV{} ($p<2$~\GeV).

\subsection*{Systematic uncertainties in electrons and photons}
The electron and photon calibration in ATLAS uses a tag-and-probe technique
in an event sample with $Z$ boson decays ($Z \to e^+ e^-$) exploiting the
constraint of the $Z$ boson mass and cross-checked with $J/\Psi$ and
$Z \to l l \gamma$ decays at low transverse energy~\cite{EGAM-2018-01,EGAM-2021-02}.
The electron and photon calorimeter response is measured in data and MC
for $\pt > 5$~\GeV.
Since the standard electron and photon calibration applies to isolated particles,
an additional uncertainty of $0.5$\% is estimated for effects in a jet environment
and to take possible differences between electrons and photons into account.
For momenta beyond the reach of the $Z \to e^+ e^-$ \insitu{} calibration,
the standard extrapolation estimate based on the detailed knowledge
of how electrons are measured with the ATLAS detector is used~\cite{PERF-2017-03,EGAM-2018-01}. %
For lower momenta ($\pt < 5$~\GeV{}), the invariant mass of neutral pions decaying into two photons is reconstructed using a minimum bias data sample.
The same energy shift between data and MC at higher momentum is found, but
to stay conservative an additional uncertainty of $1$\% is applied to low energy
electromagnetic energy deposits.

\subsection*{Modelling of various hadron types}
The most abundant particle types in jets are charged and neutral pions,
but some other particle types contribute as well.
The hadronic interactions of baryons and mesons
such as $\pi^\pm$, $p$, $\bar{p}$, $n$, $\bar{n}$, $K^\pm$, $K_L$, $K_S$, $\Lambda$ etc.
with the detector can lead to slightly different calorimeter responses~\cite{PERF-2015-05}.
The effect is small for particles with $\pt>20$~\GeV, but can be sizeable for very low \pt. %
The baseline calibration and uncertainty derived for charged pions from the \Wtaunu{} sample or a mixture
of charged hadrons dominated by pions in the minimum bias sample is applied to all hadron types, but additional uncertainties are estimated for specific hadron types based on MC simulations.

Baryons produce less neutral pions and therefore the measured
energy in the calorimeter is lower than for a pion with the same momentum.
These effects are well simulated~\cite{PERF-2015-05, Abdallah:2021gai, Adragna:2010zz}.
Studies using $\Lambda$ baryon decays in $pp$ collisions at $8$ \TeV{} show that the response of protons is about $10$\% lower
than the one for pions for momenta between $1$ and $7$~\GeV ~\cite{PERF-2015-05}. The MC simulation
describes this effect within statistical uncertainties.
Because of isospin symmetry %
neutrons behave similar to protons in the calorimeter. This is also confirmed by simulation
studies showing that protons and neutrons have the same energy deposit over the studied
kinematic range of $0.7 < p < 100$~\GeV{} \cite{PERF-2015-05}.
Possible differences between the proton and neutron response can be indirectly tested
by studying the charge exchange reaction ($\pi^- p \to \pi^0 n$ versus $\pi^+ n \to \pi^0 p$).
The calorimeter response to $\pi^+$  is about $10$\% higher than the one to $\pi^-$ at $p = 1$~\GeV, while
the response becomes the same for $p > 3$~\GeV{} \cite{PERF-2015-05}.
The charge dependence at low \pt{} is caused by the higher density of neutrons relative to protons
in the absorber material of the ATLAS electromagnetic calorimeter.
The simulation describes the data within statistical uncertainties.
Since, however, no direct constraints are available for neutrons and the above studies are statistically limited
a conservative uncertainty of $10$\% ($5$\%) is assigned for neutrons with $p<3$~\GeV{} ($p>3$~\GeV)~\cite{PERF-2015-05}.

The slightly different showering of charged kaons compared with pions is well simulated~\cite{Abdallah:2021gai}
and no additional corrections or uncertainties are taken into account.
The long-lived neutral kaon ($K_L$) lifetime is such that they are likely to reach the calorimeter before decaying
and will therefore start their hadronic shower in the calorimeter but some of them will decay before entering the calorimeter.

If they decay before the calorimeter the impinging particles are different than for a stable particle
produced in the collisions.
The uncertainties in the effect of long-lived neutral kaons on the jet energy measurement
is addressed by varying shower simulation models and detector geometries in the MC simulation.
Simulations of single particles in the ATLAS detector are used, with varied assumptions,
including different models of how particles interact with the detector and different detector geometries.

The largest deviation of the response ratio of long-lived kaons and pions
relative to the default simulation is $3\%$ and is obtained when exchanging the Fritiof model with the QGSP model.
The variations of the detector geometry are negligible.
The uncertainty is estimated conservatively to cover the full deviation.
For the region $\pt< 25$~\GeV{} a flat uncertainty of $3$\% is assumed, while for $\pt > 25$~\GeV{}
the uncertainty is parameterised as $0.05 - 0.015 \cdot \log{(\pt/\GeV)}$.

Short-lived kaons ($K_S$) decay $70$\% of the time into
$K_S \to \pi^{+} \pi^{-}$ and $30$\% into $K_S \to \pi^{0} \pi^{0}$.
Above $27$~\GeV{}, a $K_S$ on average travels over $1.47$~{\rm m} and, therefore, enters
the calorimeter before decaying. At high momentum, $K_S$ behaves similarly to $K_L$ (having a similar mass and the same charge), and both of  the particles produce showers in the calorimeter. Therefore, the same uncertainties as those for $K_L$ are applied to $K_S$ above $27$~\GeV{}.
Below $27$~\GeV{}, the $K_S$ decay mode is simulated according to the
branching ratio, and either the charged pion uncertainties
with half of the kaon momentum ($p(\pi)= 0.5 p(K_S)$) or the photon uncertainties with $p(\gamma)= 0.25 p(K_S)$ are applied.

The uncertainties in the various particle types are all applied on top of the
pion uncertainties from the \EoP{} measurements.

\subsection*{Extrapolation of the single-particle response to very high \pt}

For very high-\pt{} jets, the extrapolation of the single-particle response beyond
the measured kinematic range is important.
Two different methods are used to estimate this uncertainty.
The first method is based on variations of the hadronic shower simulations.
The second method explores the knowledge of the \JES{} response of the \pt-balance based \insitu{} methods at $2$~\TeV{}
to constrain the \JES{} derived from \EoP{} measurements at high \pt{}  by scanning various assumed values of the out-of-range particle uncertainty.

In the first method the single-particle simulations as described in Section~\ref{sec:mc} are used.
All variations are forced to agree at $\pt=175$~\GeV, where there are precise measurements
from the \Wtaunu{} data. Then the variation is studied as function of \pt{} up to $2$~\TeV.
In addition, the energy is varied by $3$\% in the hadronic barrel and by $8$\% in the hadronic endcap calorimeter based on differences seen
in the \EoP{} measurement using $\tau$-lepton decays ~\cite{JETM-2018-08}.
The biggest effect is seen for the modelling of the nuclear break up for high-\pt{} jets
using the QGS model instead of the Fritiof model. It can give a variation of up to $2\%$ at the highest \pt.
A simple parameterisation with a log-linear model, $0.02 \; \log{(\pt/175~\GeV )}$,
is used to cover the deviation.

The uncertainty in the \EoP{} measurement is assumed to remain constant beyond the last single-particle response measurement point. The extrapolation uncertainty is added in quadrature to the uncertainty in the \EoP{} measurement.

The second method studies the effect of scanning the assumed value of the out-of-range uncertainty
from $1$\% to $10$\% for particles outside of the kinematic reach of the \EoP{} measurements. Previously $10\%$ was assumed~\cite{PERF-2011-05,PERF-2015-05}.
If the out-of-range single-particle response is changed up or down, the derived jet response is correspondingly changed at high \pt.
For such a strong variation the \JES{} derived from the \EoP{} measurements is not compatible
with the \JES{} derived from \insitu{} \pt-balance methods.
The \JES{} together with its uncertainty derived from the \pt-balance \insitu{} methods
therefore provides a constraint on the single-particle out-of-range response uncertainty.
The particle spectra in the jet have large fluctuations and therefore particles at fixed \pt{} are contained
in jets with a large range in jet \pt.
The fraction of deposited energy by particles in a fixed \pt{} range increases
for increasing jet \pt. Therefore, the knowledge of the \JES{} response and of the corresponding
uncertainty at a given \pt{} can be
used to extrapolate to higher \pt.
An uncertainty of $2$\% on the single-particle out-of-range uncertainty is evaluated following this procedure, resulting in a \JES{} response uncertainty of $0.7$\% ($0.8\%$) for a jet at $\pt = 2$ ($4$)~\TeV.
This result is similar to the extrapolation uncertainty estimated from the variations in the simulation of the ATLAS detector.

In the final \JES{} uncertainty the out-of-range uncertainty from the MC model variations (first method) is used.
This is a reduction of a factor of four compared with the previously assumed out-of-range \JES{} uncertainty~\cite{PERF-2011-05,PERF-2015-05}.
It is achieved by testing the detector modelling with  \insitu{} single-particle measurements covering a wide kinematic range
and is supported through cross-checks using precise high-\pt{} jet response measurements
to constrain the high-\pt{} single-particle response outside the \pt{} range where it is measured.


%
%
%
%
%
%
\pagebreak
\section{Jet energy scale for quark- and gluon-induced jets and comparison to \pt-balance methods}
\label{sec:result}

Since the deconvolution method relies on the MC simulations to determine the \JES{} %
using the single-particle response measurements,
the \JES{} can be studied separately for quark- and gluon-induced jets.
The standard ATLAS definition for jet studies to define the jet flavour is used, i.e. the flavour of the parton with the highest energy associated to the jet determines the jet flavour.
The parton-to-jet association uses helper particles with negligible energy, but pointing in the
same direction as the partons. These so-called `ghost-particles'
are clustered together with the stable particles by the jet algorithm following the method of Ref.~\cite{Cacciari:2008gp}.
The notion of a ‘quark-initiated’ or ‘gluon-initiated’ jet is not well-defined beyond leading-order in QCD~\cite{Gras:2017jty,Komiske:2018vkc}, but it captures well the differences in the jet fragmentation. On average quark-initiated jets are narrower and have fewer constituents with a harder particle spectrum than gluon-initiated jets with the same \pt.

\begin{figure}[htp!]
\begin{center}
\subfloat[]{\includegraphics[width=0.45\linewidth]{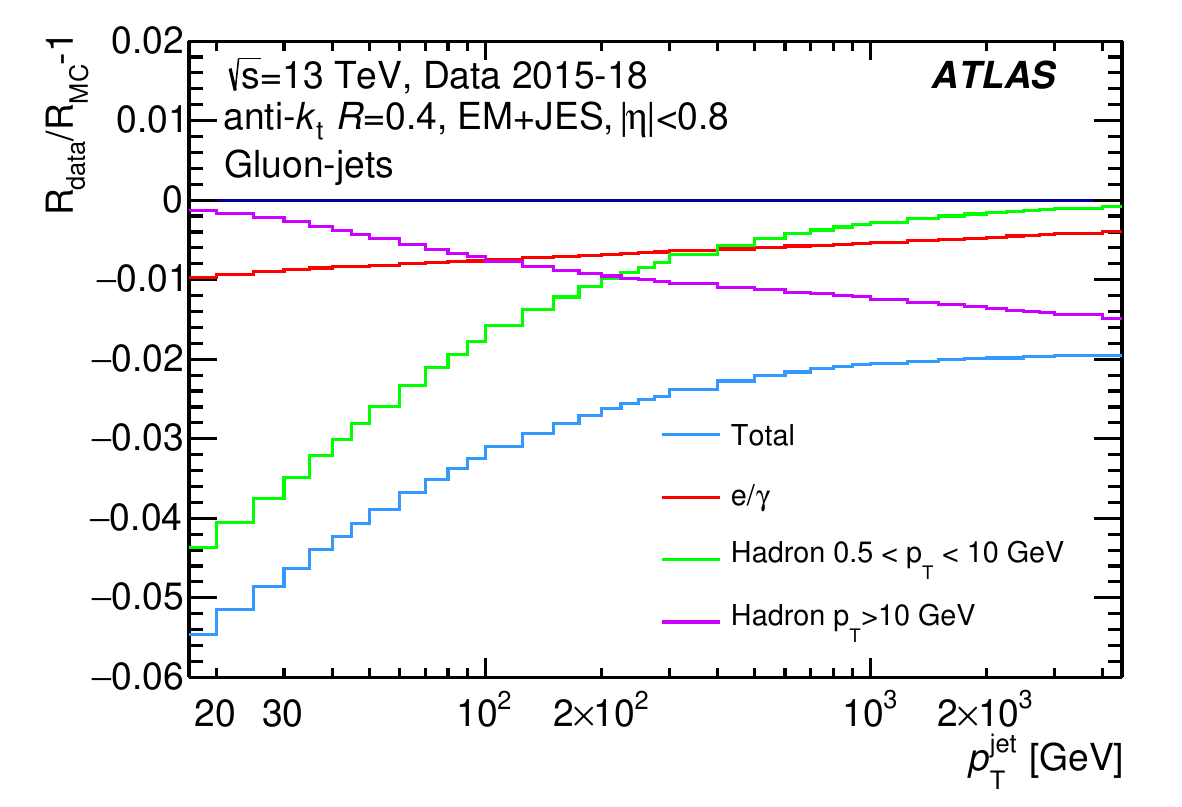}}
\subfloat[]{\includegraphics[width=0.45\linewidth]{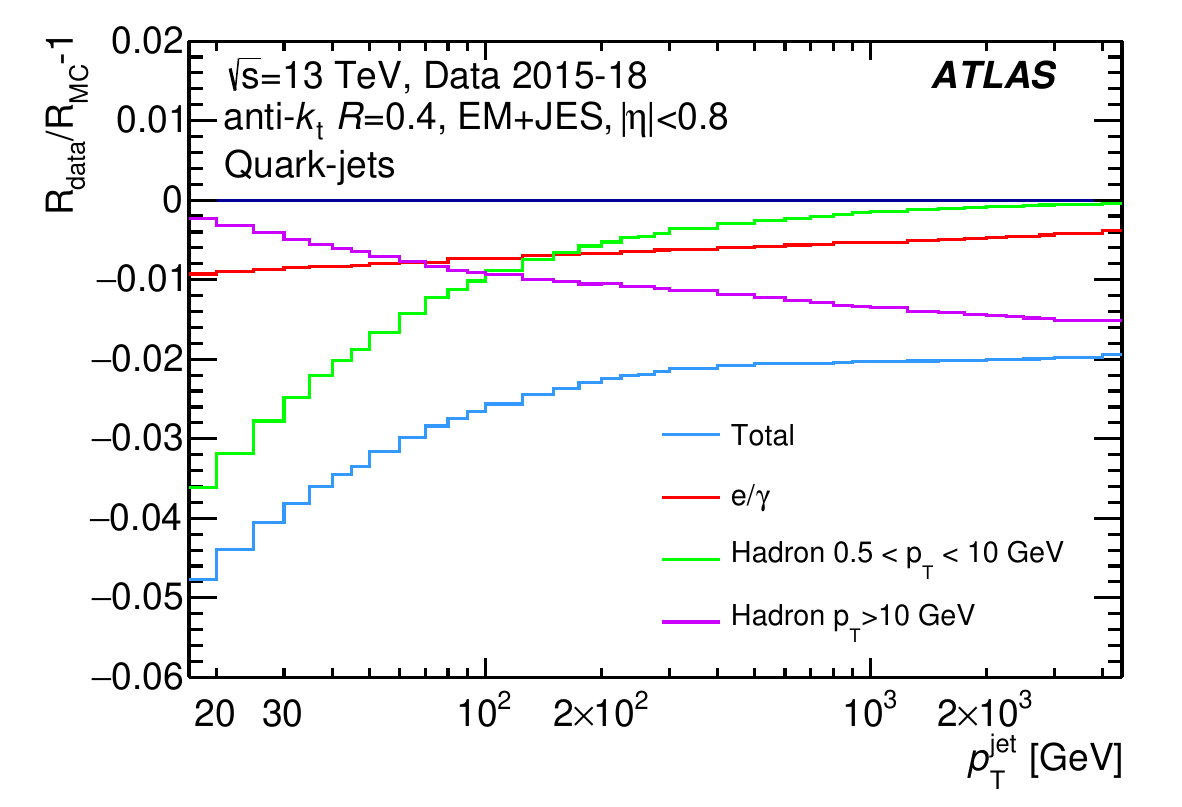}} \\
\subfloat[]{\includegraphics[width=0.45\linewidth]{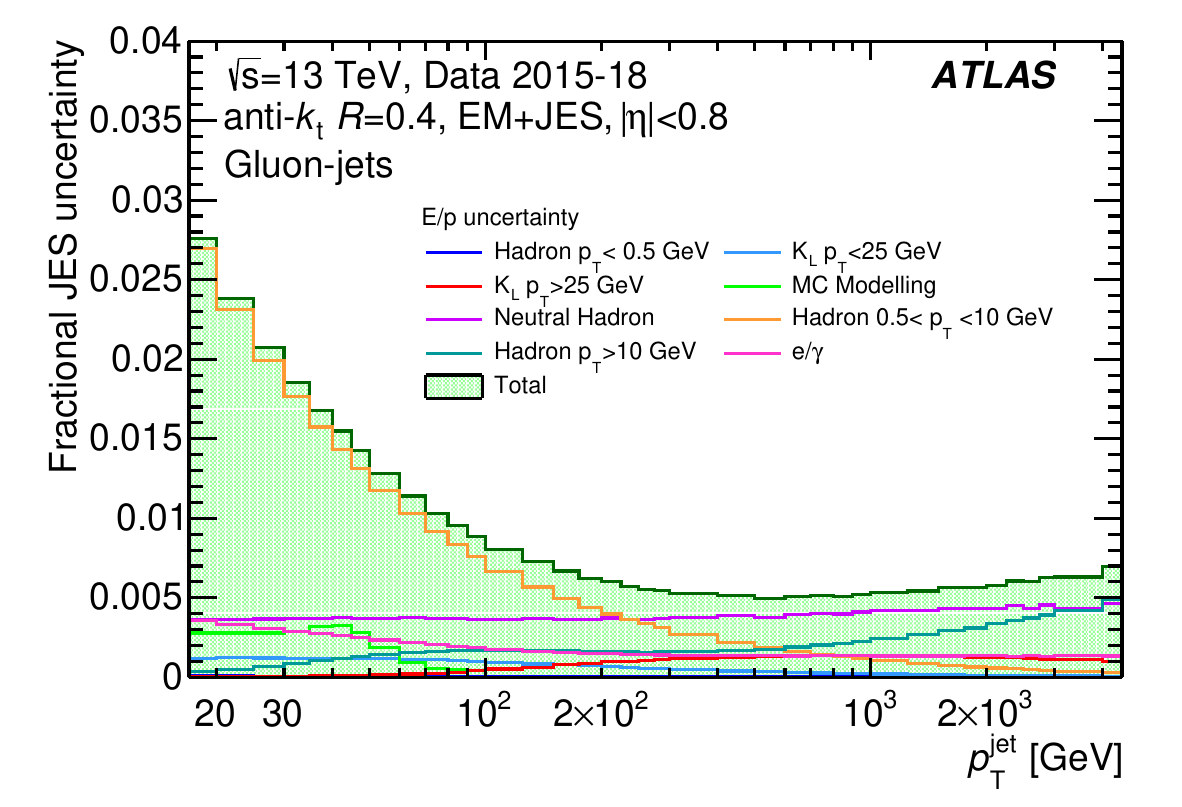}}
\subfloat[]{\includegraphics[width=0.45\linewidth]{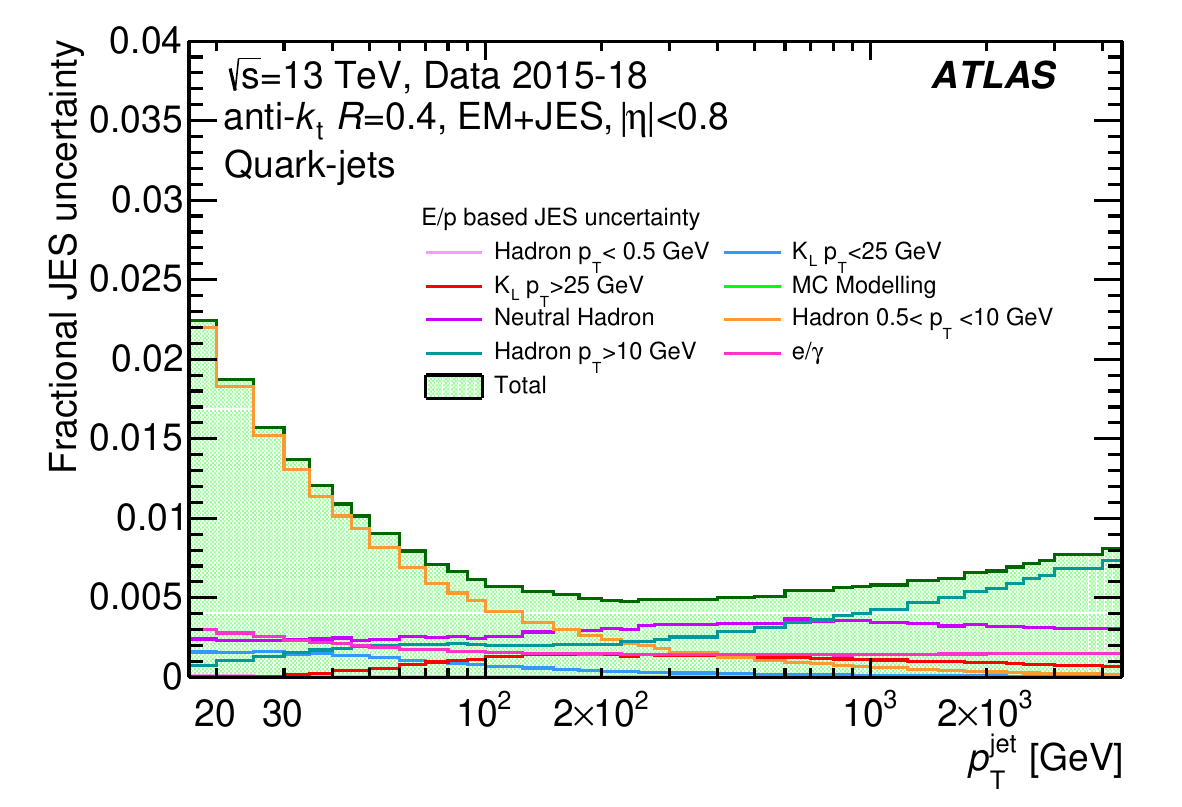}}
\end{center}
\caption{
Data-to-MC jet response shift for (a) gluon-induced and (b) quark-induced jets.
The total correction and the individual shifts due to the $e/\gamma$ correction,
the single-hadron response measurement at low-\pt{} and the single-pion response at high-\pt{} are shown.
The horizontal dark blue line indicates no correction.
Fractional jet energy scale uncertainty derived from the \EoP{} measurements as a function of the jet \pt{}
for (c) gluon-induced and (d) quark-induced jets. The individual systematic uncertainties and the resulting total uncertainty are shown.
}
\label{fig:quarkgluons}
\end{figure}

Figures~\ref{fig:quarkgluons}(a) and \ref{fig:quarkgluons}(b) show the response shifts separately for gluon- and quark-induced jets
for the main three single-particle measurements individually and the resulting total correction.

The individual corrections are obtained by selecting either electrons and photons ($e/\gamma$)
or hadrons in exclusive kinematic domains at low \pt{} (\EoP{} in the minimum bias data sets)
or at high \pt{} (\EoP{} in the \Wtaunu{} data sets). The total correction is obtained using all particles.
The jet response in data is lower than the one in MC. Therefore the jet response
shift (${\rm R}_{\rm Data} / {\rm R}_{\rm MC}-1$)
is negative for both of the types of jets.
The total jet response correction is about $-5\%$ at low-\pt and  $-2\%$ at high-\pt{} for quark-induced jets.
The response correction for gluon-induced jets is slightly more negative than the one for quark-induced jets.
Gluon-induced jet have a softer fragmentation, resulting in a higher fraction of the total jet energy contained
in low-\pt{} particles. The lower overall response shift is caused by the low-\pt{} measurements in the minimum bias data set.
The correction for electrons and photons induces a shift of $-1$\% at low \pt{} that gets a bit reduced towards higher \pt, i.e.
for $\pt = 4000$~\GeV{} it is $- 0.4$\%.
The single-hadron response measurement predicts a response shift of
-$4$\% at low \pt{} and no
shift at high \pt. The single-pion measurement from the $\tau$-lepton decays predicts
a response shift of -$0.2$\% at low \pt{} increasing to -$1.5$\% for $\pt = 4000$~\GeV.

The individual uncertainties of the \JES{} derived from the single-particle response measurements
are shown in Figures~\ref{fig:quarkgluons}(c) and \ref{fig:quarkgluons}(d) as a function of the jet \pt.
The $e/\gamma$ uncertainties contribute about $0.1$\% to $0.2$\% almost independent of \pt.
The dominating uncertainty at low \pt{} is the uncertainty
due to tracks that do not leave clusters in the calorimeter. This effect dominates the total
uncertainty up to about $\pt = 200$~\GeV. In the range of $200 < \pt < 2000$~\GeV{} the largest single uncertainty is due to
the calorimeter response to neutral hadrons. Beyond this \pt{} range the extrapolation uncertainty is largest.
The uncertainties are slightly larger for gluon- than for quark-induced jets,
since the uncertainty for low momentum particles have a higher impact on gluon-induced jets due to their softer fragmentation pattern.

\begin{figure}[htp!]
\begin{center}
\includegraphics[width=0.8\linewidth]{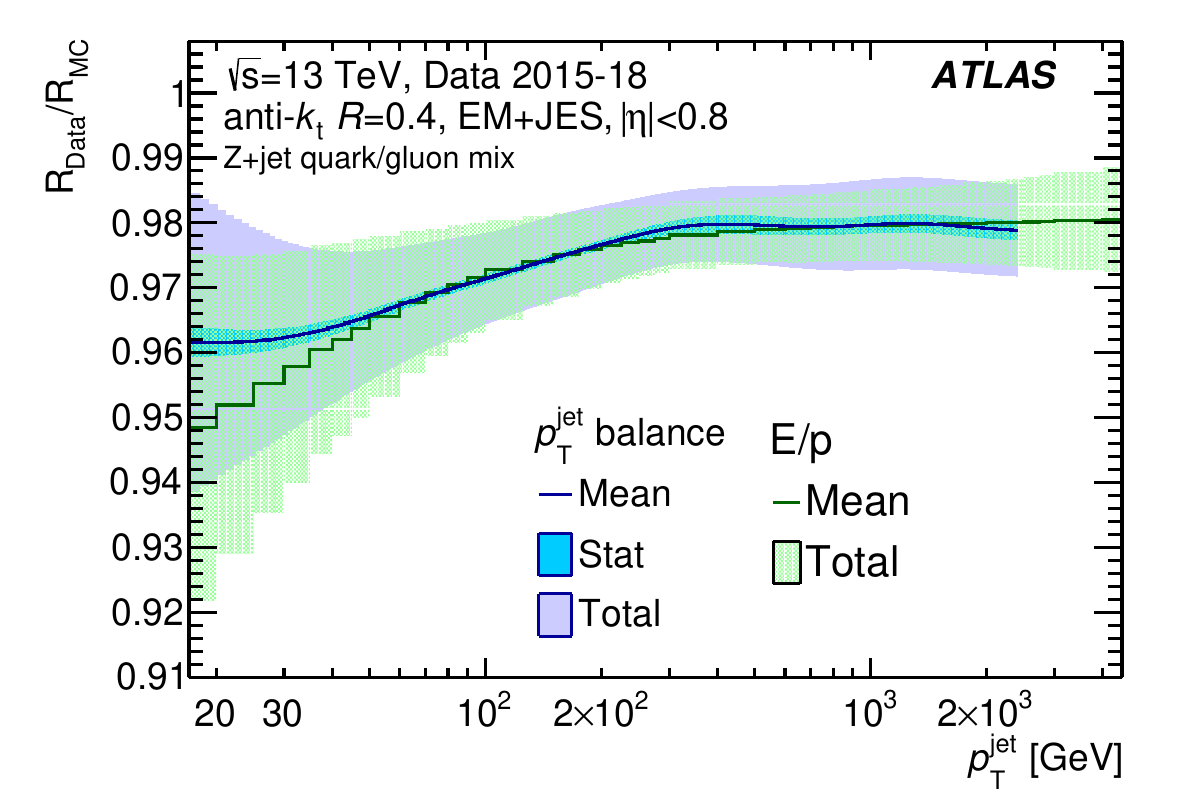}
\end{center}
\caption{
Data-to-MC \JES{} response ratio derived from the \pt-balance based \insitu{} methods and the one from the single-particle response measurements (\EoP) as a function of the jet \pt. The quark/gluon fractions as expected in the $Z$+jet process are used. The bands indicate the statistical and total uncertainties.
}
\label{fig:JES_ep_vs_ptbalance_quarkgluon}
\end{figure}

The \JES{} response and uncertainty as calculated from the single-particle measurement
is shown in \Fig{\ref{fig:JES_ep_vs_ptbalance_quarkgluon}}
as a function of the jet \pt{} together with the \JES{} from the
\pt-balance based \insitu{} techniques.
Here the jet response is calculated with the relative amount of
quark- and gluon-induced jets (quark/gluon mixture) chosen
to correspond to the expectation for jets in $Z/\gamma$+jets events.
With this quark/gluon mixture a good agreement with the \pt-balance based method is obtained.
However, any other quark/gluon composition would also give agreement within uncertainties.
The response shift between data and MC simulation of about -$2$\% at high \pt{}
obtained with two completely independent methods is very similar and well within
the quoted uncertainties. The drop of the response ratio towards smaller \pt{} is also reproduced within uncertainties.
The uncertainty from the single-particle response measurements
is smaller than the one from the $Z/\gamma$+jet \insitu{} methods for about $\pt > 200$~\GeV.
The single-particle-based uncertainty smoothly and slowly increases towards the highest \pt{} values.
At $2000$~\GeV{} the \pt-balance based \JES{} uncertainty is $0.8\%$ while the one from the single particles is $0.6$\%.
The single-particle response measurement can be used for all jet \pt, i.e. also in the region where the \pt-balance based methods do not have kinematic reach.
%


\pagebreak
\section{Combination of single-particle response and \pt-balance methods}
\label{sec:combination}

To achieve the highest precision determination of the jet energy scale corrections and uncertainties a combination of the results from single-particle response measurements and the \pt-balance based \insitu{} techniques is performed.
The combination is performed using a spline-based procedure,
that has also been employed in e.g.\ Refs.~\cite{PERF-2011-03,PERF-2012-01,PERF-2016-04,JETM-2018-02,PERF-2014-02,JETM-2018-05} and references therein.
Either linear or quadratic splines are used to interpolate between the various points of each \insitu{} measurement and between the points of the \EoP{} result.
They are used to determine the contributions of each of these inputs in fine \pt{} bins.

An uncertainty propagation procedure, based on pseudo-experiments (propagating the full information of the input uncertainties with their correlations), allows to determine the covariance matrix of the various contributions in a given fine \pt{} bin.
A weighted average of these contributions, based on a $\chi^2$ minimisation, is then performed. The weights for the various \insitu{} \pt-balance based and \EoP{} inputs, computed taking into account the corresponding statistical and systematic uncertainties, should not be biased by the bin-size choices that were made for each of them. For this reason, the determination of the averaging weights is performed using \insitu{} \pt-balance and \EoP{} inputs that are first integrated into a common large binning, before performing the spline-based interpolation. The original \insitu{} \pt-balance and \EoP{} binning is employed afterwards, when performing the actual averaging using these weights.

While yielding a combined result that can be used for the jet energy calibration, it also allows for the evaluation of the level of agreement between the inputs.
Each of the input uncertainties~(i.e. nuisance parameters) is propagated through the combination procedure, by shifting the impacted inputs by one standard deviation of that uncertainty, re-doing the full averaging procedure and comparing the result with the nominal one.
The fit quality is evaluated based on the  $\chi^2/{\rm Ndof}$, where Ndof is the number of degrees of freedom.
In case of some tensions between the inputs, identified through a $\chi^2/{\rm Ndof} > 1$ in a given bin,
the propagated uncertainties are %
increased by a factor $\sqrt{ \chi^2/{\rm Ndof} }$, following the recommendations of e.g. Ref.~\cite{Zyla:2020zbs}.
Finally, a Gaussian-kernel smoothing is applied coherently to the combination result and to its uncertainties.

\begin{figure}[htp!]
\begin{center}
\subfloat[]{\includegraphics[width=0.85\linewidth]{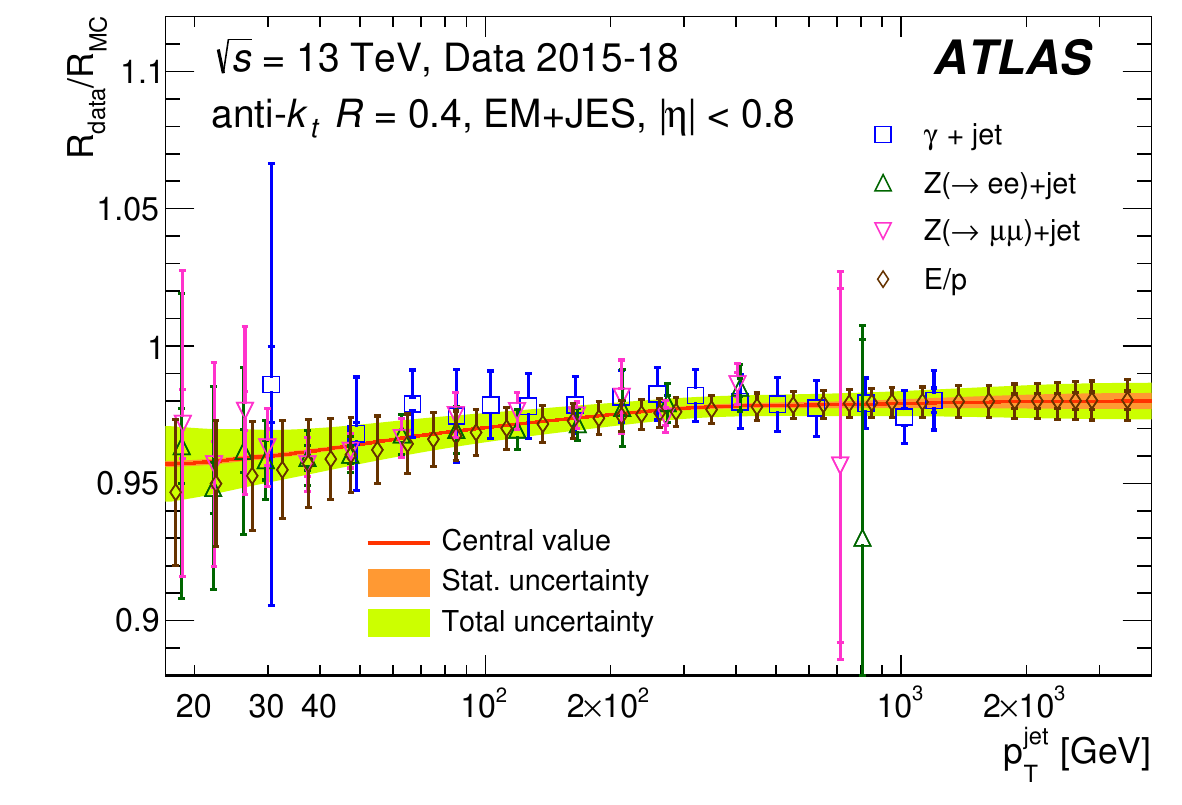}} \\
\subfloat[]{\includegraphics[width=0.45\linewidth]{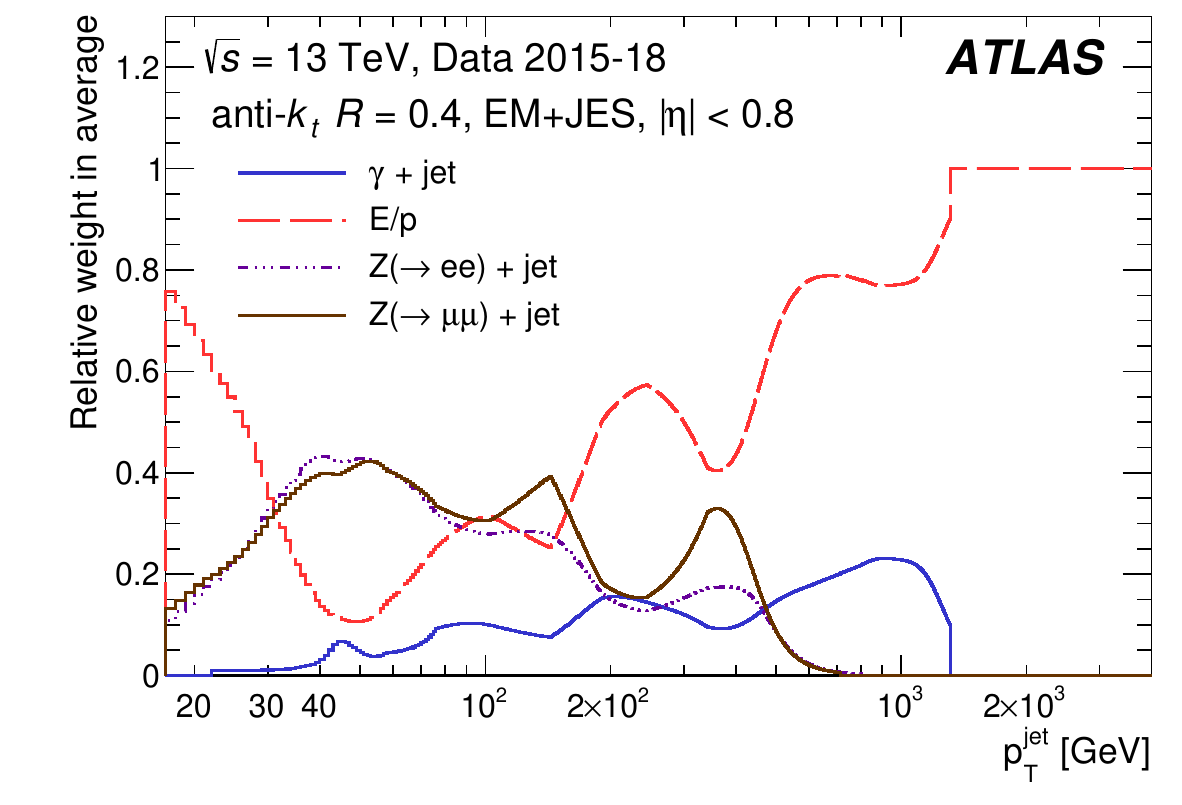}}
\subfloat[]{\includegraphics[width=0.45\linewidth]{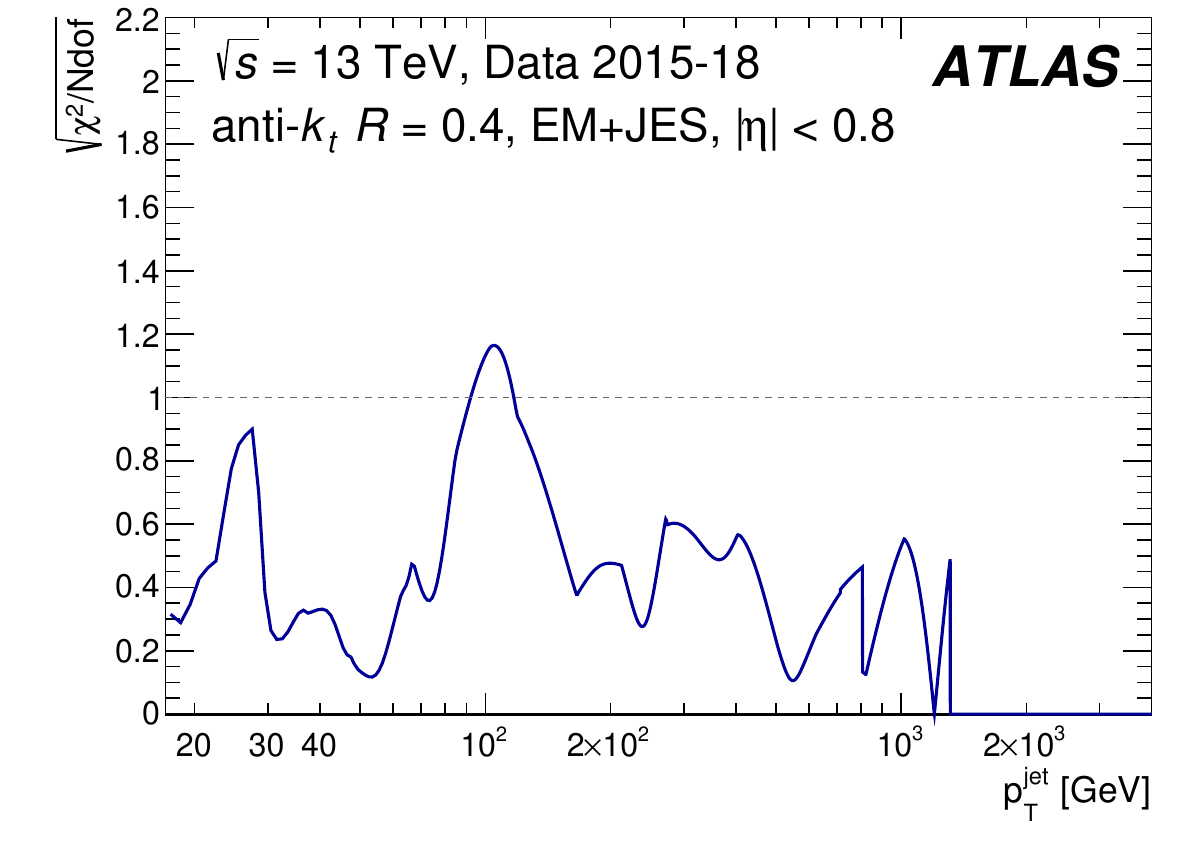}}
\end{center}
\caption{
Combination of the \insitu{} \EoP{} and  \pt-balance ($\gamma+{\rm jet}$ and $Z+{\rm jet}$)~\cite{ATLAS:2023tyv} results for the \EMTopo{} jet collection.
(a) Data-to-MC jet response ratio
as a function of the jet \pt. The points indicate the various measurements with their statistical (inner bars) and systematic uncertainties added in quadrature (outer bars). The line indicates the nominal combination result, while bands indicate its statistical (inner bands) and total uncertainty (outer band) respectively.
(b) The relative weights of the various input measurements used in the combination and (c) the $\sqrt{ \chi^2/{\rm Ndof} }$ as a function of the jet \pt.
}
\label{fig:WeightsChi2_NoMJB_EMtopo_TSpline1}
\end{figure}

The baseline result of this study, with the \EMTopo{} jet collection, employs the precise \EoP{} result down to \pt{} values of $17~\GeV$.
The combination result, with the corresponding \insitu{} \pt-balance and \EoP{} inputs, the relative combination weights and the $\sqrt{ \chi^2/{\rm Ndof} }$, as a function of the jet \pt, are shown in Figure~\ref{fig:WeightsChi2_NoMJB_EMtopo_TSpline1}.
In contrast to previous combinations ~\cite{PERF-2011-03,PERF-2012-01,PERF-2016-04,JETM-2018-02,PERF-2014-02,JETM-2018-05},
the results from the multi-jet balance (MJB) used to extend the \pt-range of the $Z/\gamma$+jet results
are not used anymore, since the high-\pt{} region is now covered by the \EoP{} measurements and which are more precise than the MJB method.
The MJB results are in excellent agreement with the \EoP{} measurements and while adding them does not give an uncertainty improvement, it would add a large number of small additional systematic uncertainty components. Thus the MJB is not included going forward.

The \EoP{} result has a significant weight for the full \pt{} range
and is the most significant contribution across the majority of the range where it is used.
For instance,
the weight of the \EoP{} method increases with \pt{} from $10$\% at $40$~\GeV{} to $75$\% at $1$~\TeV{}.
At very low \pt{} like  $20$~\GeV{} the weight is $70$\%, thus helping to reduce the large uncertainty from the $Z+{\rm jets}$ modelling.
The \insitu{} \pt-balance based and the \EoP{} inputs are also in good agreement over the full \pt{} range
with a $\chi^2$/Ndof around $0.5$ in most of the \pt-range. At $100$~\GeV{} the $\chi^2$/Ndof is around $1$.

In \RunTwo{} most ATLAS physics results are based on the \PFlow{} jet reconstruction algorithm~\cite{PERF-2015-09} where
tracks and calorimeter-clusters are combined to improve the jet resolution and to mitigate pile-up effects.
The \PFlow{} algorithm benefits from the fact that charged hadrons at low \pt{} can be better measured with the tracking detector than with the calorimeter. Moreover,
the tracks can be better assigned to the hard scattering vertex and therefore the rejection of pile-up contributions is easier.
At high \pt, however, charged hadrons can be better measured with the calorimeter and the contribution from tracks becomes minimal.
The response of jets based on the \PFlow{} algorithm therefore tends to the one from \EMTopo{} jets towards high-\pt~\cite{JETM-2018-05}.

In the ATLAS \PFlow{} algorithm a decision is made for each track whether the associated energy deposit in the calorimeter
can be accurately identified such that it can be subtracted and the track measurement used~\cite{JETM-2018-05}.
This is dependent on the track \pt{} and the energy density around the shower in the calorimeter as in dense environments this
identification becomes difficult. To correct for non-compensation effects, for the particles where the subtraction is not performed
a fraction of the track energy is still included in the jet. Importantly, for particles where this subtraction is not performed, the
shifted calorimeter contributions can be used to evaluate the shift in the jet energy scale of \PFlow{} jets.
Since \PFlow{} jets also use tracks to reconstruct the jet energy, the jet response corrections take the
reduced calorimeter energy fraction into account.
For jets with \pt $>$ $264$~\GeV{} the fraction of the jet constituents where the subtraction is performed is below $5$\%,
and thus the impact of track corrections is expected to be small.
Therefore the single-particle technique can be applied to derive the jet energy response corrections and uncertainties.
To avoid the region where this approximation might not be so accurate the combination of \insitu{} \pt-balance techniques
and single-particle response measurements is only performed above $300$~\GeV.
The modelling of the charged-particle energy fraction in \PFlow{} jets, i.e. the sum of the energy in tracks with respect to the jet \pt,
was measured previously~\cite{PERF-2015-09,PERF-2011-03} and is well modelled.
Therefore no additional uncertainty is added.

\begin{figure}[htp!]
\begin{center}
\subfloat[]{\includegraphics[width=0.85\linewidth]{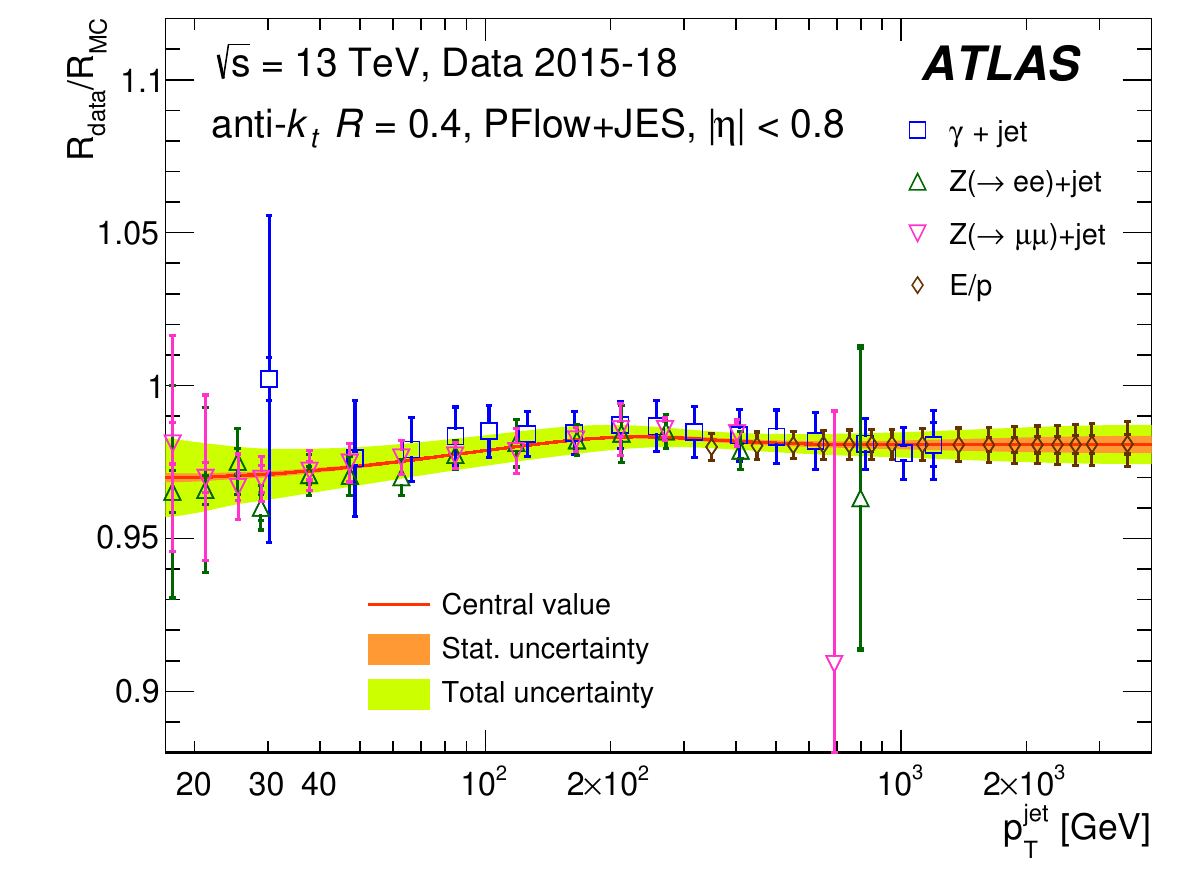}} \\
\subfloat[]{\includegraphics[width=0.45\linewidth]{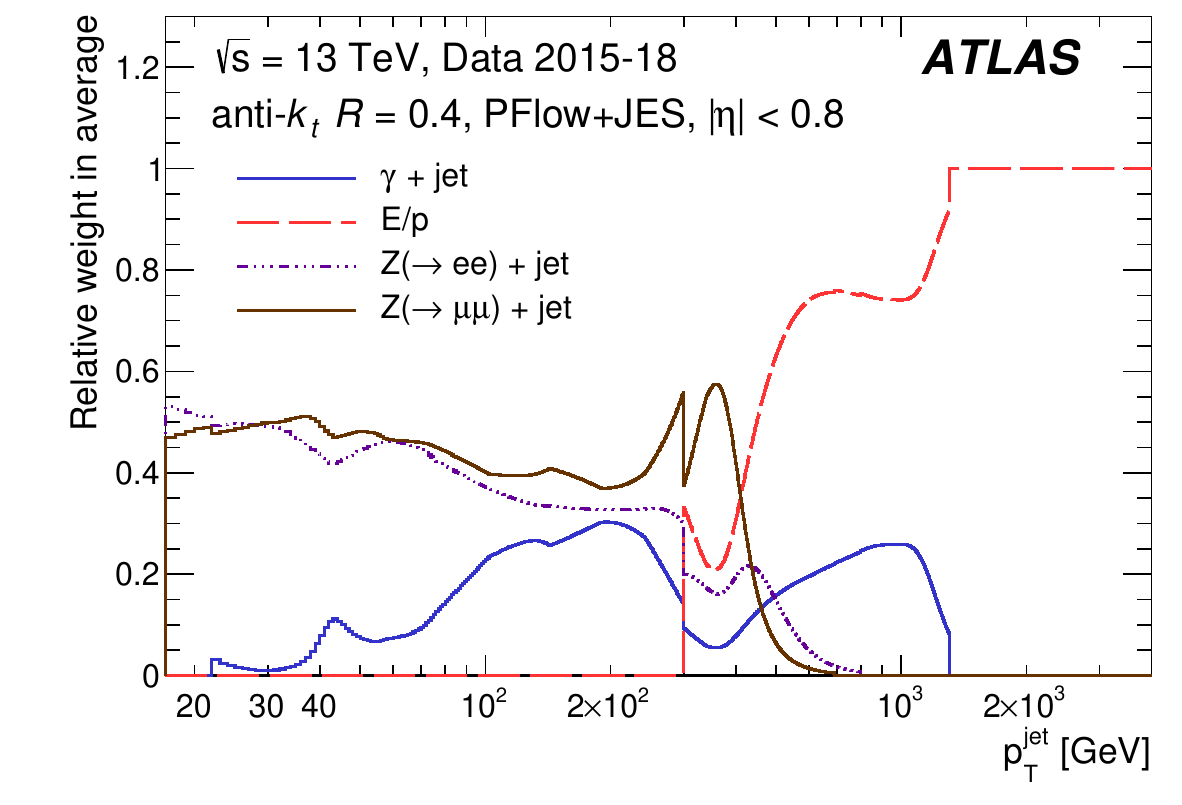}}
\subfloat[]{\includegraphics[width=0.45\linewidth]{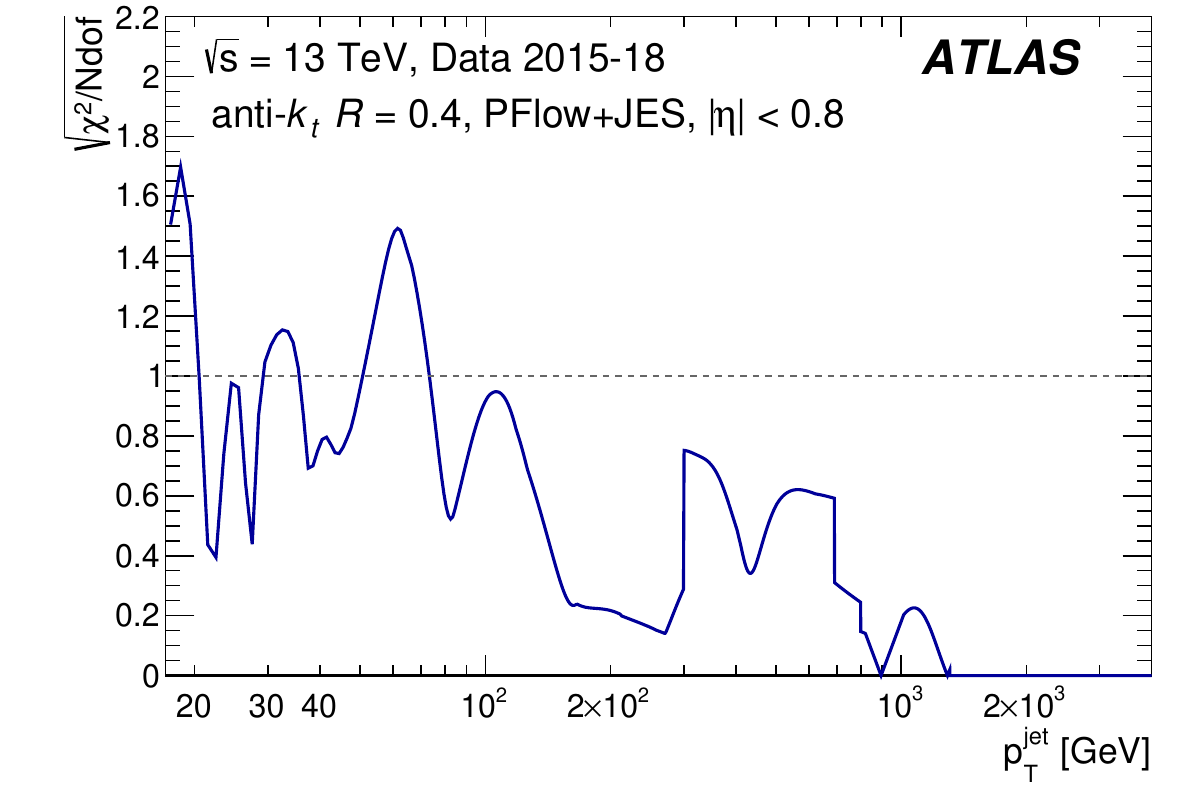}}
\end{center}
\caption{
Combination of the \insitu{} \EoP{} and  \pt-balance ($\gamma+{\rm jet}$ and $Z+{\rm jet}$)~\cite{ATLAS:2023tyv} results for the \PFlow{} jet collection.
(a) Data to MC jet response ratio
as a function of the jet \pt. The points indicate the various measurements with their statistical (inner bars) and systematic uncertainties added in quadrature (outer bars). The line indicates the nominal combination result, while bands indicate its statistical (inner bands) and total uncertainty (outer band) respectively.
(b) The relative weights of the various input measurements used in the combination and (c) the $\sqrt{ \chi^2/{\rm Ndof} }$ as a function of the jet \pt.
}
\label{fig:WeightsChi2_NoMJB_pFlow_TSpline1}
\end{figure}

The combination result for \PFlow{} jets, with the corresponding \insitu{} \pt-balance based and \EoP{} inputs, the relative combination weights and the $\sqrt{ \chi^2/{\rm Ndof} }$, as a function of the jet \pt, are shown in Figure~\ref{fig:WeightsChi2_NoMJB_pFlow_TSpline1}.
Here also, the \EoP{} result has a relatively important weight on average, for the full \pt{} range where it is used.
At $300$~\GeV{} it contributes about $20$~\%, at $1$~\TeV{} almost $80$\% and $100$~\% for $\pt > 1500$~\GeV.
The \insitu{} \pt-balance based and \EoP{} inputs are also in good agreement in the full \pt{} range where they overlap.
The $\chi^2$/Ndof is below $0.6$ over the full \pt{} range where the combination is performed.


\pagebreak

%
%


%
\pagebreak
\section{Comparison to previous results}
\label{sec:comparison}

\begin{figure}[htp!]
\begin{center}
\subfloat[]{\includegraphics[width=0.49\linewidth]{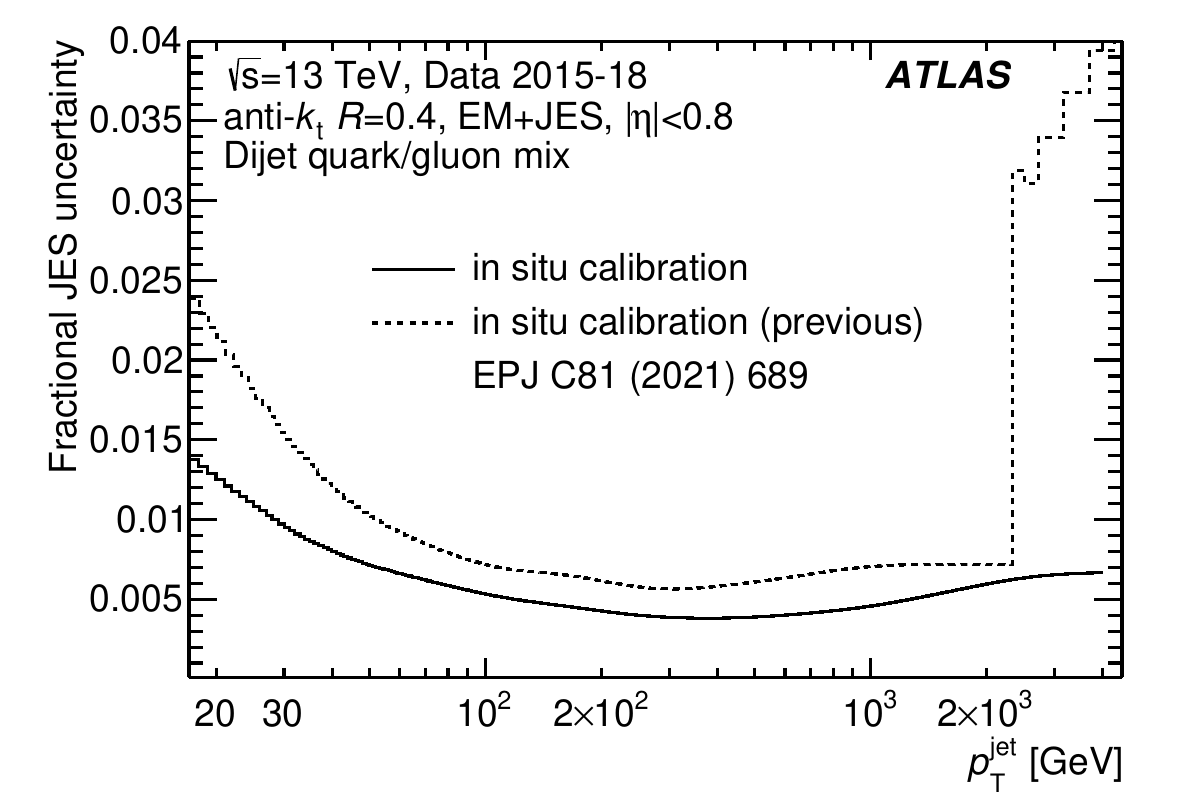}}
\subfloat[]{\includegraphics[width=0.49\linewidth]{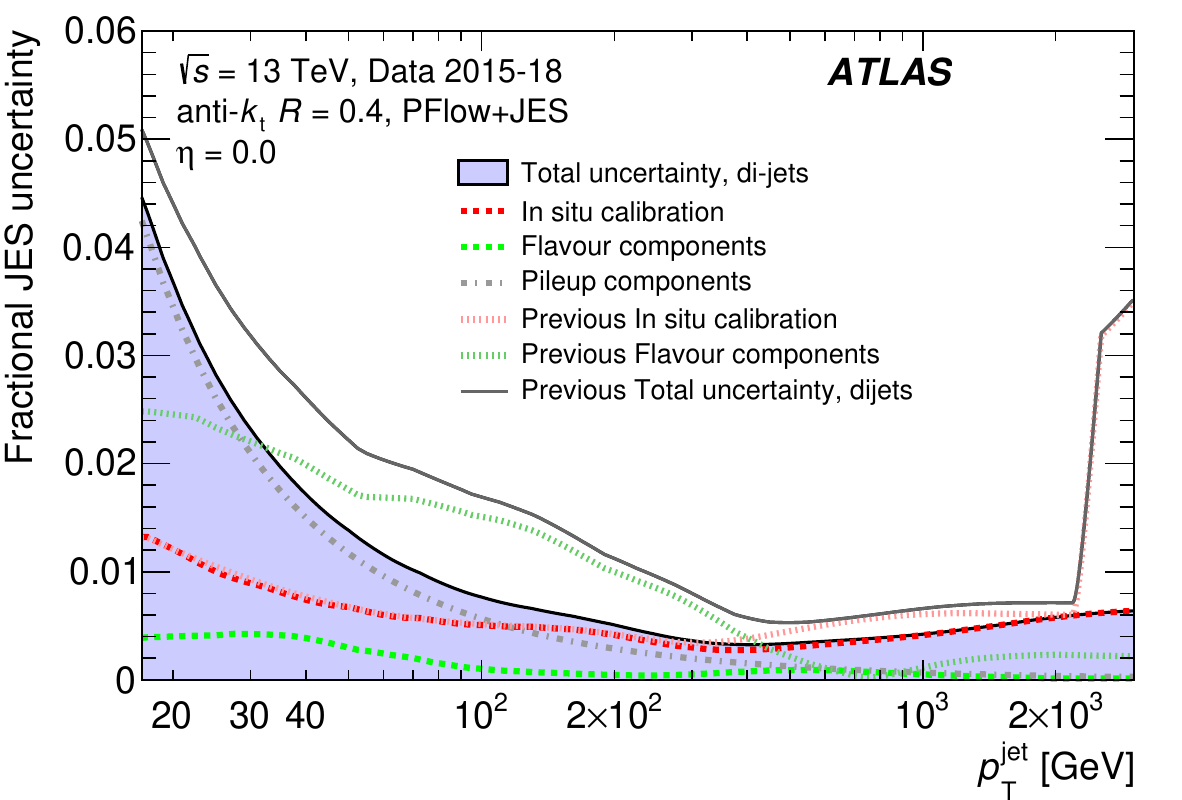}}
\end{center}
\caption{
(a) Fractional \JES{} uncertainty in the data-to-MC response ratio
from the \insitu{} calibration using the new (solid) and the previously recommended \JES{} uncertainty  \cite{JETM-2018-05}  (dashed) for \EMTopo{} jets as a function of the jet \pt.
(b)
Total fractional \JES{} for \PFlow{} jets in the most central pseudo-rapidity region as a function of the jet \pt.
In addition to the \JES{} from the \insitu{} calibration (red dashed line) the additional uncertainties due to the pile-up (grey dot-dashed line) and the jet flavour components added in quadrature (green dashed line) are shown.
The blue area shows the quadrature sum of all the uncertainties. For comparison the uncertainties as previously recommended for physics analysis are also given: \insitu{} calibration (red dotted line), flavour components (green dotted line) and total (solid grey line). The improved flavour uncertainties are taken from Refs.~\cite{ATL-PHYS-PUB-2022-021, ATLAS:2024png}.
}
\label{fig:jetcomparison}
\end{figure}

The main part of the jet calibration is based on \insitu{} techniques that assess the difference
between data and Monte Carlo simulation of the calorimeter response to jets (\JES{} from \insitu{} calibration).
The reduction in the jet calibration uncertainties for \EMTopo{} jets is illustrated
in Figure~\ref{fig:jetcomparison}(a) that compares the uncertainty obtained by the combination
of the two \insitu{} techniques discussed in this paper
with the previous ones used in ATLAS measurements~\cite{JETM-2018-05}.
A significant improvement is seen for all values of jet \pt.
The large improvement by a factor of $3$ to $4$
above roughly $2.5$~\TeV{} is mainly due to the larger kinematic range covered in the \EoP{} measurements in the \Wtaunu{} sample
and the additional studies constraining the single-particle uncertainties beyond the measured range (see Section~\ref{sec:method}).

However, in addition to the \insitu{} calibration \JES{} uncertainties, there are also additional uncertainties to account for the dependence on the varying pile-up conditions
(pile-up components) and on the flavour of the parton initiating the jet~\cite{PERF-2014-02, PERF-2015-09, PERF-2016-04, JETM-2018-05, ATLAS:2023tyv} (flavour components).
Recently, significant progress has recently been made on the understanding
of the calorimeter response to different jet flavours~\cite{ATL-PHYS-PUB-2022-021, ATLAS:2024png}.
It was observed that the jet response difference between \Pythia{} and \herwig{} can be explained
by the differences in the baryon fraction of the particles constituting the jet. \Sherpa{} (\cite{Gleisberg:2008ta,Hoeche:2009rj}) using the cluster model~\cite{Webber:1983if, Winter:2003tt} was in agreement with  \herwig{}
but agreed with \Pythia{} after a tuning of the baryon fraction to LEP data~\cite{Chahal:2022rid}.
By calibrating the jet response in \herwig{}~\cite{Corcella:2000bw,Bellm:2015jjp} to the one in  \Pythia{} a large reduction of the flavour uncertainty was obtained.
As a result the jet flavour uncertainty was reduced to a negligible level.
Moreover, the flavour uncertainty was split into
several components by comparing MC samples with different hadronisation models (string vs cluster model),
different parton showers and different modelling
of the hard scattering process (matrix element).

It is now interesting to see how both recent improvements compare to what is presently used
in ATLAS physics analysis.
Figure~\ref{fig:jetcomparison}(b) shows the previous total \JES{} uncertainty~\cite{JETM-2018-05}
and the one presented here for a quark/gluon mix in an inclusive dijet sample for \PFlow{} jets.
A significant improvement is seen for all jet \pt.
Also shown are the additional uncertainties due to pile-up and to the jet flavour.
The flavour uncertainty is very small (below $0.4$\% for $\pt < 100$~\GeV{} and negligibly small for $\pt > 100$~\GeV).
The pile-up uncertainty is the largest uncertainty for jets below $\pt<100$~\GeV.
For higher \pt{} the uncertainties related to the \insitu{} \JES{} calibration are largest.
The total \JES{} uncertainty amounts to about $4.0$\% at $\pt=20$~\GeV,
below $1$\%  at $\pt = 100$~\GeV{}, $0.4$\% at $\pt = 300$~\GeV{} and $0.6$\% at $\pt=3$~\TeV.
The uncertainty at low \pt{} is dominated by the pile-up component
but recently new pile-up subtraction techniques have been developed to significantly reduce these uncertainties~\cite{ATLAS:2023tyv}.
The flavour uncertainties are significantly reduced by the updated treatment and the combination of \insitu{} techniques
with the newly derived corrections from single-particle measurements provide a significant reduction in this component above $400$~\GeV.

%


%

\pagebreak
\section{Jets with various radius parameters}
\label{sec:rscan}

The deconvolution method allows for the derivation of the data-to-MC response corrections for jets reconstructed with
radius parameters of the \AKT{} jet algorithm different from the ATLAS default value ($\Rjet = 0.4$).
In particular, studies of jet substructure for both small-\Rjet{} and large-\Rjet{} jets are interesting.
In this section results are presented for \Rjet-values ranging from $0.2$ to $1.0$.
The \JES{} can be studied with or without applying clustering algorithms to remove soft jet constituents (jet grooming).
\begin{figure}[htp!]
\begin{center}
\subfloat[]{\includegraphics[width=0.49\linewidth]{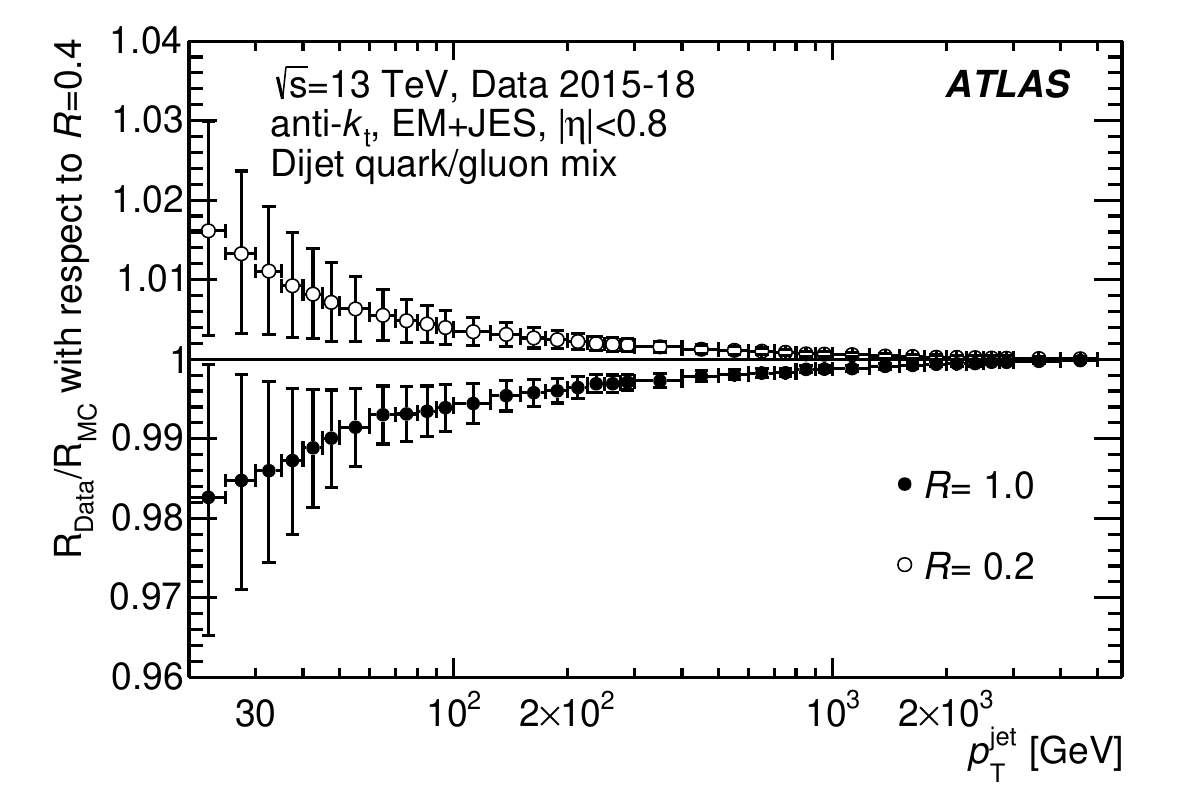}}
\subfloat[]{\includegraphics[width=0.49\linewidth]{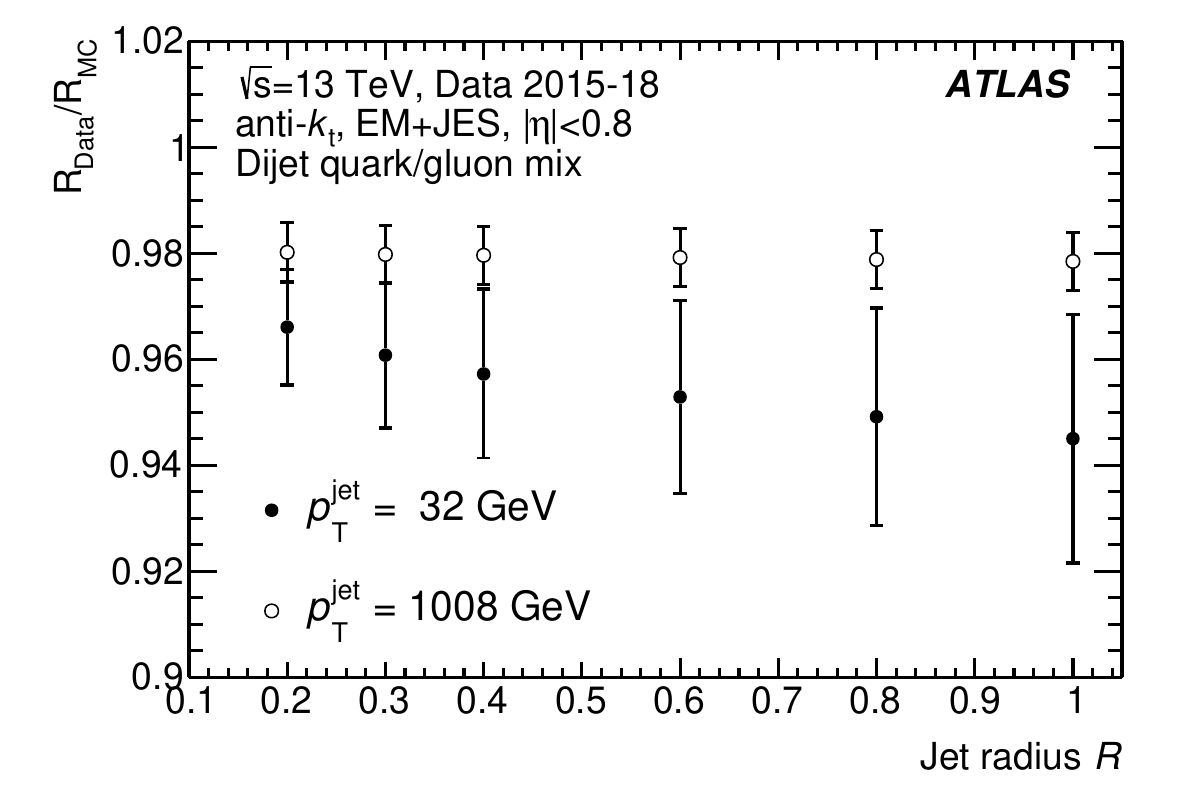}}
\end{center}
\caption{
(a) Data-to-MC \JES{} response ratio derived from the single-particle response measurements for jets with
$\Rjet = 0.2$ and  $\Rjet = 1.0$ normalised to the one of jets with $\Rjet = 0.4$
as a function of the jet \pt.  The horizontal line at unity is drawn for better visibility.
(b) Data-to-MC \JES{} response ratio for jets with
$\pt = 32$~\GeV{} and $\pt = 1008$~\GeV{} as a function of the jet radius.
The error bars respresent the quadrature sum of all the individual  uncertainties.
The flavour composition as in the inclusive dijet MC simulation is used.
}
\label{fig:Rscanresponse}
\end{figure}
Figure~\ref{fig:Rscanresponse}(a) shows the data-to-MC jet response correction together with their uncertainties for jets with radius parameters $\Rjet = 0.2$ and $\Rjet = 1.0$
divided by the one of $\Rjet= 0.4$ jets as a function of the jet \pt.
Since the systematic uncertainties are correlated
between the various \Rjet{} values, the presentation of the results relative to the
standard jets with $\Rjet = 0.4$ helps to better visualise the dependence
of the jet response correction on the jet radius.
The small radius jets have the smallest jet response correction and the largest radius jets the largest.
At low \pt{} differences between the \Rjet{} values up to about $3$\% are observed. At large \pt{}
the differences are below $1$\%, but are still significant.
For $1$~\TeV{} jets the response correction for $\Rjet = 1.0$ ($\Rjet = 0.2$) is $0.15$\% ($0.06$\%) lower (higher) than the
one for $\Rjet = 0.4$ jets. For jets with $\pt > 2$~\TeV{} the response shifts are within the
systematic uncertainties.

Figure~\ref{fig:Rscanresponse}(b) shows the data-to-MC response correction as a function
of the jet radius parameter for jets with $\pt=32$~\GeV{} and with $1008$~\GeV.\footnote{A low-\pt{} and a high-\pt{} value is displayed as example. The quoted \pt{} values correspond to the beginning of the chosen \pt{} bin which is more relevant than the bin centre since in most physics processes the \pt{} distribution falls steeply.}
Here no ratio with respect to $\Rjet = 0.4$
is taken and the response correction can be seen together with the uncertainty
for each jet radius.
In this presentation the systematic uncertainties are strongly correlated across the \Rjet{} bins.
At $\pt=32$~\GeV{} the  $\Rjet = 0.2$ jets have a jet response shift of $-3.5$\% while the largest jet with  $\Rjet =1.0$
has a response shift of $-5.5$\%. For $\pt = 1008$~\GeV{} a smaller variation
is observed with the jet radius. When taking the correlation between the jet radii into account, this
variation is also significant.

\begin{figure}[htp!]
\begin{center}
\subfloat[]{\includegraphics[width=0.49\linewidth]{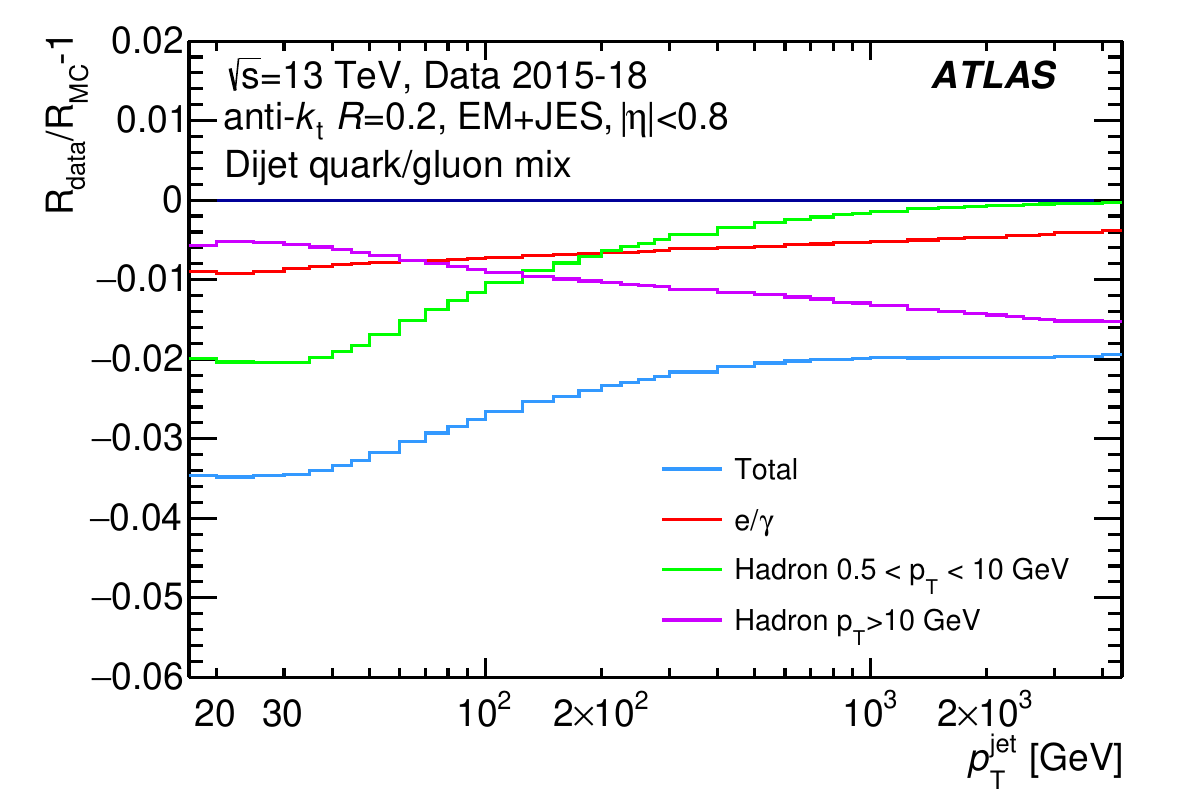}}
\subfloat[]{\includegraphics[width=0.49\linewidth]{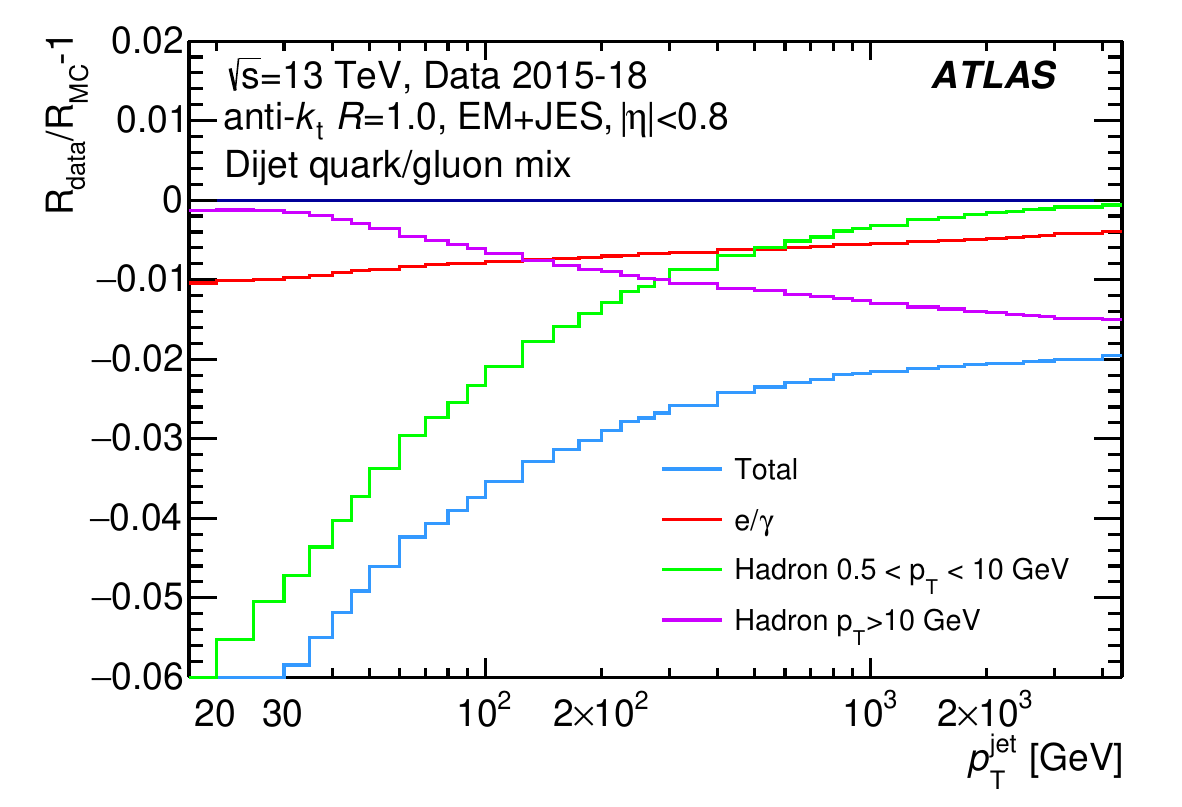}} \\
\subfloat[]{\includegraphics[width=0.49\linewidth]{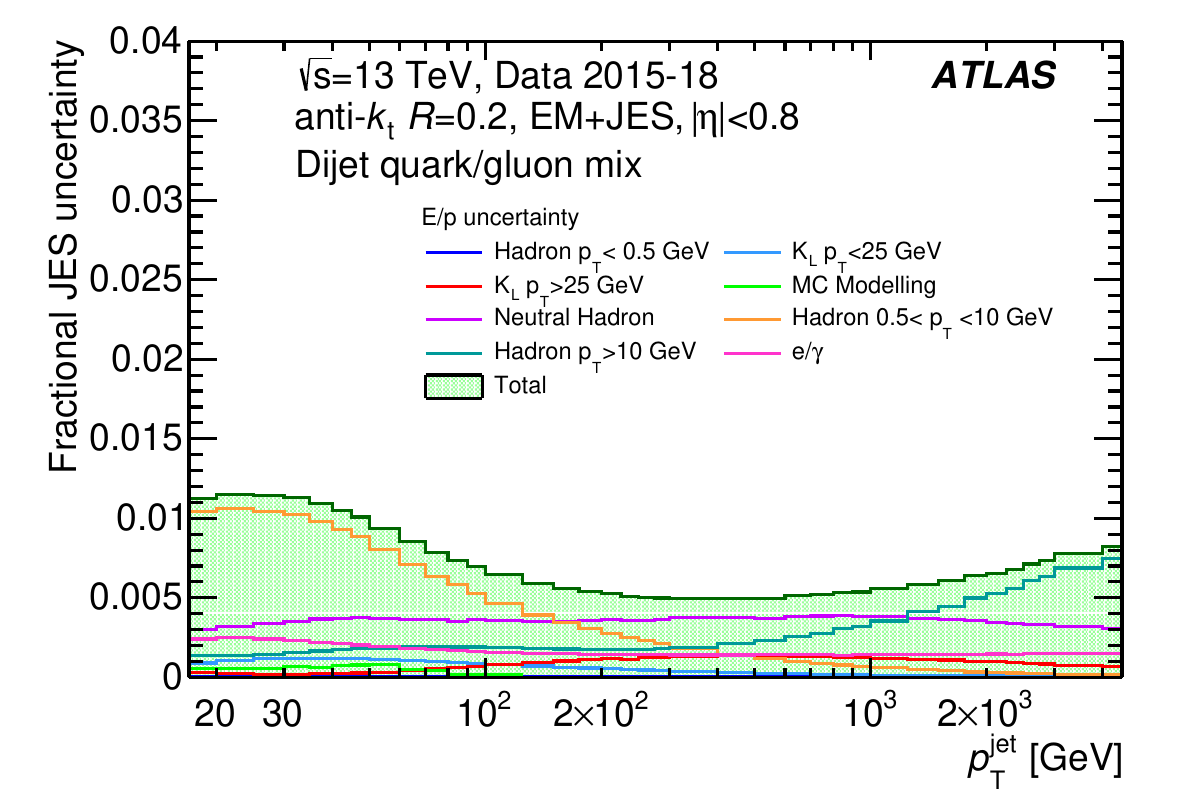}}
\subfloat[]{\includegraphics[width=0.49\linewidth]{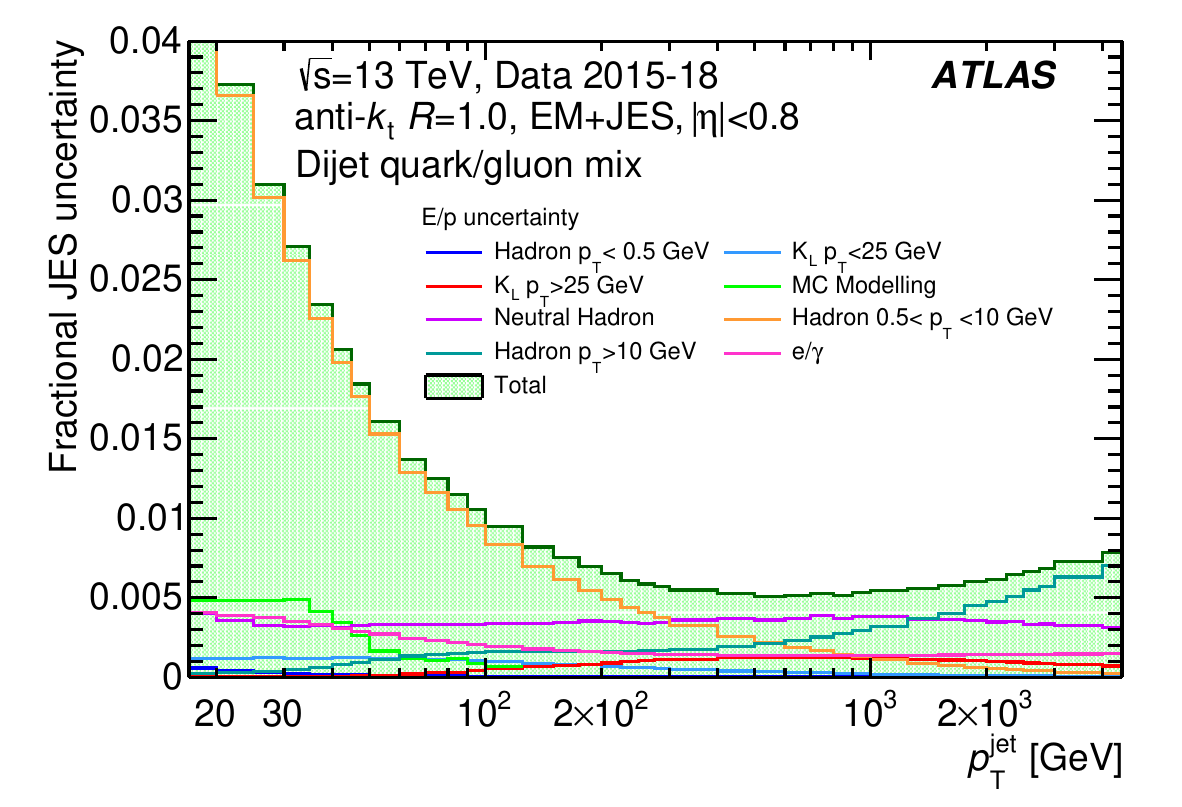}}
\end{center}
\caption{
Data-to-MC jet response correction derived from the \EoP{} measurements as a function of the jet \pt{} with (a) $\Rjet = 0.2$ and (b) $\Rjet = 1.0$. Shown are the total correction together with the individual shifts due to the $e/\gamma$ correction, the single-hadron response measurement at low-\pt{} and the single-pion response from $\tau$-lepton decays at high-\pt. The horizontal black line indicates no correction.
Fractional \JES{}  uncertainty derived from the \EoP{} measurements as a function of the jet \pt{} for jets with (c) $\Rjet = 0.2$ and with (d) $\Rjet = 1.0$.
The uncertainties are merged in classes for better visibility.
The flavour composition as in the inclusive dijet MC simulation is used.
}
\label{fig:Rscanuncertainties}
\end{figure}

Figures~\ref{fig:Rscanuncertainties}(a) and \ref{fig:Rscanuncertainties}(b) show the response correction due to the various
\EoP{} measurements for the smallest ($\Rjet{} = 0.2$) and the largest ($\Rjet = 1$) \Rjet-values as examples.
The response shifts due to the measurements of electrons and photons and of high-\pt{} pions do not vary with the \Rjet-value. However, the response shift due to the low-\pt single-hadron response measurement in minimum bias collisions has a strong variation with the \Rjet-value that is responsible for the overall response shift.
The response shift due to low-\pt{} particles changes from $-2$\% to -$5\%$ for $\pt=30$~\GeV.
Since the response shift is driven by low-\pt{} particles measured in minimum bias events, the overall response
correction does not depend on the \Rjet-value at high-\pt.
This indicates that the response difference between small-\Rjet{} and large-\Rjet{} jets, for quark- and gluon-induced jets,
is mainly driven by low-\pt{} particles.
This is in-line with ATLAS jet substructure measurements showing that the energy density is highest
close to the jet axis and that only a relatively small energy fraction is located in the outer parts~\cite{STDM-2010-10,STDM-2017-16}.
Therefore, when enlarging the jet radius the energy fraction in the low-\pt particles is
increased thus leading to an increase of the overall uncertainty.

Figures~\ref{fig:Rscanuncertainties}(c) and \ref{fig:Rscanuncertainties}(d) show the uncertainty for the two \Rjet{} values discussed above. The uncertainty due to the \EoP{} measurement in minimum bias events remains dominant, but its magnitude is smallest for small-\Rjet{} values and largest for large-\Rjet{} values. At $\pt = 30$~\GeV{} it is about $1$\% for $\Rjet = 0.2$ jets but is $3$\% for
$\Rjet = 1.0$ jets. At the highest \pt{} the variation is much less.
The other uncertainties do not dependent on the \Rjet-values.

These results are obtained for the standard jets formed by strongly interacting particles in dijet events.
Large-\Rjet{} jets are used to identify the decay products from boosted particles such as $W/Z$ bosons or top quarks
produced at large \pt. To reduce the increased sensitivity of large-\Rjet{} jets to soft radiation, particles
produced by the underlying event collisions or in pile-up $pp$ interactions,
grooming algorithms are applied that remove soft jet constituents
based on a clustering algorithm running on the jet constituents.
For an overview on this topic see Refs.~\cite{Larkoski:2017jix,Kogler:2018hem}.

In ATLAS the best performance of boosted particle reconstruction~\cite{JETM-2018-06} is obtained for
jets with $\Rjet  = 1.0$ that are used together with the soft drop clustering algorithm~\cite{Larkoski:2014wba}
with the radial distance parameter set to $\beta=1.0$ and the soft-drop threshold set to  $z=0.1$.

Without further requirements on substructure variables to increase the fraction of jets
with a two- or three-prong structure, the data-to-MC jet response correction is not changed
for an inclusive jet sample. In an event sample with boosted top quarks produced by \Zprime{} boson decays to top quark pairs
(see Section~\ref{sec:mc}), the data-to-MC jet response correction with and without jet grooming
is almost the same. For $\pt > 200$~\GeV{} they agree within  $0.1$\%.
While grooming with the soft drop algorithm does not significantly modify the data-to-MC jet response correction,
the origin of the jet does have an impact.
For $\pt > 200$~\GeV{} the response in the dijet sample is about $0.2$\% higher than in the \Zprime{} sample independent of \pt.

In summary, these studies show that differences of internal jet structure between inclusive jets and jets from boosted top-quark decays also lead to differences in the jet response correction, while the application of the soft-drop grooming algorithm with $\beta=1.0$ and $z=0.1$ has a negligible impact on the response of both the types of jets.


\pagebreak
\section{Conclusion}
\label{sec:conclusion}

The jet energy scale correction together with the corresponding uncertainties has been derived from
measurements of the calorimeter response to single particles of the ATLAS detector operating at the LHC.
In particular, the measurements of isolated pions produced in $\tau$-lepton decays
have significantly increased the kinematic range where the calorimeter response to single pions
has been measured \insitu. As a result the jet energy scale correction uncertainty is significantly improved, reaching
$2.5$\% at $\pt=20$~\GeV, $0.5$\% at $\pt=300$~\GeV{} and $0.7$\% at $\pt=4$~\TeV.

The results obtained from the single-particle measurements and the \pt-balance based method
are in excellent agreement with the single-particle-based method having an improved accuracy.
Since the uncertainties of the two measurements are uncorrelated,
the combination of the two methods significantly improves the jet energy scale measurement with respect to earlier estimates.
While low-\pt{} jets are still dominated by uncertainties related to pile-up, at $\pt = 300$~\GeV{}
and for \PFlow{} jets
the final \JES{} uncertainty amounts to $0.3$\% and increases to $0.6$\% at $\pt=3$~\TeV.

Furthermore, the single-particle response method allows for the determination of the
data-to-MC jet response correction for any jet collection definition, such as for
variations of the jet radius parameter. Differences between jets of different radii are significant.
While inclusive jets and jets from boosted top-quark decays have a jet response correction within $0.2$\%
for $\pt>200$~\GeV, the application of the soft-drop grooming algorithm with $\beta=1.0$ and $z=0.1$ has a negligible impact on the response of both types of jets.

This analysis demonstrates a fundamental understanding of the jet energy measurement and
a good description of the physics processes in the calorimeter by the MC simulation.
The data-to-MC jet response shift as measured with the \insitu{} \pt-balance based methods
observed since the beginning of the data taking at the LHC
can be shown to be due to the deficiencies of the MC simulation to describe
the response of the ATLAS calorimeter to single particles.
The jet response shift is primarily driven by the single hadron response modelling, but also
has significant contributions from the modelling of electromagnetic processes.

The improvements in the jet energy calibration presented in this paper pave the way
for precision jet measurements with the ATLAS detector.


%

%
%
%
%
%
\section*{Acknowledgements}
%

%

%
%

%
%

We thank CERN for the very successful operation of the LHC and its injectors, as well as the support staff at
CERN and at our institutions worldwide without whom ATLAS could not be operated efficiently.

The crucial computing support from all WLCG partners is acknowledged gratefully, in particular from CERN, the ATLAS Tier-1 facilities at TRIUMF/SFU (Canada), NDGF (Denmark, Norway, Sweden), CC-IN2P3 (France), KIT/GridKA (Germany), INFN-CNAF (Italy), NL-T1 (Netherlands), PIC (Spain), RAL (UK) and BNL (USA), the Tier-2 facilities worldwide and large non-WLCG resource providers. Major contributors of computing resources are listed in Ref.~\cite{ATL-SOFT-PUB-2025-001}.

We gratefully acknowledge the support of ANPCyT, Argentina; YerPhI, Armenia; ARC, Australia; BMWFW and FWF, Austria; ANAS, Azerbaijan; CNPq and FAPESP, Brazil; NSERC, NRC and CFI, Canada; CERN; ANID, Chile; CAS, MOST and NSFC, China; Minciencias, Colombia; MEYS CR, Czech Republic; DNRF and DNSRC, Denmark; IN2P3-CNRS and CEA-DRF/IRFU, France; SRNSFG, Georgia; BMBF, HGF and MPG, Germany; GSRI, Greece; RGC and Hong Kong SAR, China; ISF and Benoziyo Center, Israel; INFN, Italy; MEXT and JSPS, Japan; CNRST, Morocco; NWO, Netherlands; RCN, Norway; MNiSW, Poland; FCT, Portugal; MNE/IFA, Romania; MSTDI, Serbia; MSSR, Slovakia; ARIS and MVZI, Slovenia; DSI/NRF, South Africa; MICIU/AEI, Spain; SRC and Wallenberg Foundation, Sweden; SERI, SNSF and Cantons of Bern and Geneva, Switzerland; NSTC, Taipei; TENMAK, T\"urkiye; STFC/UKRI, United Kingdom; DOE and NSF, United States of America.

Individual groups and members have received support from BCKDF, CANARIE, CRC and DRAC, Canada; CERN-CZ, FORTE and PRIMUS, Czech Republic; COST, ERC, ERDF, Horizon 2020, ICSC-NextGenerationEU and Marie Sk{\l}odowska-Curie Actions, European Union; Investissements d'Avenir Labex, Investissements d'Avenir Idex and ANR, France; DFG and AvH Foundation, Germany; Herakleitos, Thales and Aristeia programmes co-financed by EU-ESF and the Greek NSRF, Greece; BSF-NSF and MINERVA, Israel; NCN and NAWA, Poland; La Caixa Banking Foundation, CERCA Programme Generalitat de Catalunya and PROMETEO and GenT Programmes Generalitat Valenciana, Spain; G\"{o}ran Gustafssons Stiftelse, Sweden; The Royal Society and Leverhulme Trust, United Kingdom.

In addition, individual members wish to acknowledge support from Armenia: Yerevan Physics Institute (FAPERJ); CERN: European Organization for Nuclear Research (CERN PJAS); Chile: Agencia Nacional de Investigaci\'on y Desarrollo (FONDECYT 1230812, FONDECYT 1230987, FONDECYT 1240864); China: Chinese Ministry of Science and Technology (MOST-2023YFA1605700, MOST-2023YFA1609300), National Natural Science Foundation of China (NSFC - 12175119, NSFC 12275265, NSFC-12075060); Czech Republic: Czech Science Foundation (GACR - 24-11373S), Ministry of Education Youth and Sports (FORTE CZ.02.01.01/00/22\_008/0004632), PRIMUS Research Programme (PRIMUS/21/SCI/017); EU: H2020 European Research Council (ERC - 101002463); European Union: European Research Council (ERC - 948254, ERC 101089007), Horizon 2020 Framework Programme (MUCCA - CHIST-ERA-19-XAI-00), European Union, Future Artificial Intelligence Research (FAIR-NextGenerationEU PE00000013), Italian Center for High Performance Computing, Big Data and Quantum Computing (ICSC, NextGenerationEU); France: Agence Nationale de la Recherche (ANR-20-CE31-0013, ANR-21-CE31-0013, ANR-21-CE31-0022, ANR-22-EDIR-0002), Investissements d'Avenir Labex (ANR-11-LABX-0012); Germany: Baden-Württemberg Stiftung (BW Stiftung-Postdoc Eliteprogramme), Deutsche Forschungsgemeinschaft (DFG - 469666862, DFG - CR 312/5-2); Italy: Istituto Nazionale di Fisica Nucleare (ICSC, NextGenerationEU), Ministero dell'Università e della Ricerca (PRIN - 20223N7F8K - PNRR M4.C2.1.1); Japan: Japan Society for the Promotion of Science (JSPS KAKENHI JP22H01227, JSPS KAKENHI JP22H04944, JSPS KAKENHI JP22KK0227, JSPS KAKENHI JP23KK0245); Netherlands: Netherlands Organisation for Scientific Research (NWO Veni 2020 - VI.Veni.202.179); Norway: Research Council of Norway (RCN-314472); Poland: Ministry of Science and Higher Education (IDUB AGH, POB8, D4 no 9722), Polish National Agency for Academic Exchange (PPN/PPO/2020/1/00002/U/00001), Polish National Science Centre (NCN 2021/42/E/ST2/00350, NCN OPUS nr 2022/47/B/ST2/03059, NCN UMO-2019/34/E/ST2/00393, NCN \& H2020 MSCA 945339, UMO-2020/37/B/ST2/01043, UMO-2021/40/C/ST2/00187, UMO-2022/47/O/ST2/00148, UMO-2023/49/B/ST2/04085, UMO-2023/51/B/ST2/00920); Slovenia: Slovenian Research Agency (ARIS grant J1-3010); Spain: Generalitat Valenciana (Artemisa, FEDER, IDIFEDER/2018/048), Ministry of Science and Innovation (MCIN \& NextGenEU PCI2022-135018-2, MICIN \& FEDER PID2021-125273NB, RYC2019-028510-I, RYC2020-030254-I, RYC2021-031273-I, RYC2022-038164-I), PROMETEO and GenT Programmes Generalitat Valenciana (CIDEGENT/2019/027); Sweden: Carl Trygger Foundation (Carl Trygger Foundation CTS 22:2312), Swedish Research Council (Swedish Research Council 2023-04654, VR 2018-00482, VR 2021-03651, VR 2022-03845, VR 2022-04683, VR 2023-03403), Knut and Alice Wallenberg Foundation (KAW 2018.0157, KAW 2018.0458, KAW 2019.0447, KAW 2022.0358); Switzerland: Swiss National Science Foundation (SNSF - PCEFP2\_194658); United Kingdom: Leverhulme Trust (Leverhulme Trust RPG-2020-004), Royal Society (NIF-R1-231091); United States of America: U.S. Department of Energy (ECA DE-AC02-76SF00515), Neubauer Family Foundation.

%
%


%
%
%
%
%
%

%
%

%
%
%
\printbibliography

\clearpage
 
\begin{flushleft}
\hypersetup{urlcolor=black}
{\Large The ATLAS Collaboration}

\bigskip

\AtlasOrcid[0000-0002-6665-4934]{G.~Aad}$^\textrm{\scriptsize 104}$,
\AtlasOrcid[0000-0001-7616-1554]{E.~Aakvaag}$^\textrm{\scriptsize 17}$,
\AtlasOrcid[0000-0002-5888-2734]{B.~Abbott}$^\textrm{\scriptsize 123}$,
\AtlasOrcid[0000-0002-0287-5869]{S.~Abdelhameed}$^\textrm{\scriptsize 119a}$,
\AtlasOrcid[0000-0002-1002-1652]{K.~Abeling}$^\textrm{\scriptsize 56}$,
\AtlasOrcid[0000-0001-5763-2760]{N.J.~Abicht}$^\textrm{\scriptsize 50}$,
\AtlasOrcid[0000-0002-8496-9294]{S.H.~Abidi}$^\textrm{\scriptsize 30}$,
\AtlasOrcid[0009-0003-6578-220X]{M.~Aboelela}$^\textrm{\scriptsize 45}$,
\AtlasOrcid[0000-0002-9987-2292]{A.~Aboulhorma}$^\textrm{\scriptsize 36e}$,
\AtlasOrcid[0000-0001-5329-6640]{H.~Abramowicz}$^\textrm{\scriptsize 155}$
% ...
\bigskip
\\

% ...
$^{17}$Department for Physics and Technology, University of Bergen, Bergen; Norway.\\
% ...
$^{30}$Physics Department, Brookhaven National Laboratory, Upton NY; United States of America.\\
% ...
$^{36}$$^{(a)}$Facult\'e des Sciences Ain Chock, Universit\'e Hassan II de Casablanca;$^{(b)}$Facult\'{e} des Sciences, Universit\'{e} Ibn-Tofail, K\'{e}nitra;$^{(c)}$Facult\'e des Sciences Semlalia, Universit\'e Cadi Ayyad, LPHEA-Marrakech;$^{(d)}$LPMR, Facult\'e des Sciences, Universit\'e Mohamed Premier, Oujda;$^{(e)}$Facult\'e des sciences, Universit\'e Mohammed V, Rabat;$^{(f)}$Institute of Applied Physics, Mohammed VI Polytechnic University, Ben Guerir; Morocco.\\
% ...
$^{45}$Physics Department, Southern Methodist University, Dallas TX; United States of America.\\
% ...
$^{50}$Fakult\"{a}t Physik , Technische Universit{\"a}t Dortmund, Dortmund; Germany.\\
% ...
$^{56}$II. Physikalisches Institut, Georg-August-Universit\"{a}t G\"ottingen, G\"ottingen; Germany.\\
% ...
$^{104}$CPPM, Aix-Marseille Universit\'e, CNRS/IN2P3, Marseille; France.\\
% ...
$^{119}$$^{(a)}$New York University Abu Dhabi, Abu Dhabi;$^{(b)}$United Arab Emirates University, Al Ain; United Arab Emirates.\\
% ...
$^{123}$Homer L. Dodge Department of Physics and Astronomy, University of Oklahoma, Norman OK; United States of America.\\
% ...
$^{155}$Raymond and Beverly Sackler School of Physics and Astronomy, Tel Aviv University, Tel Aviv; Israel.\\
% ...
$^{a}$ Also Affiliated with an institute covered by a cooperation agreement with CERN.\\
$^{b}$ Also at An-Najah National University, Nablus; Palestine.\\
$^{c}$ Also at Borough of Manhattan Community College, City University of New York, New York NY; United States of America.\\
$^{d}$ Also at Center for High Energy Physics, Peking University; China.\\
$^{e}$ Also at Center for Interdisciplinary Research and Innovation (CIRI-AUTH), Thessaloniki; Greece.\\
$^{f}$ Also at CERN, Geneva; Switzerland.\\
% ...

\end{flushleft}


%

%
%
%
%

%

%
%
%
%

\end{document}